\documentclass[%
 reprint,
 amsmath,amssymb,
 aps,
]{revtex4-2}

\usepackage{graphicx}
\usepackage{dcolumn}
\usepackage{amsmath}
\usepackage{amssymb}
\usepackage{bm}
\usepackage{mathptmx}
\usepackage[utf8]{inputenc}
\usepackage[T1]{fontenc}
\usepackage{mathptmx}
\usepackage{xcolor}
\usepackage{subfig}
\usepackage[format=plain,labelfont=up,textfont=up]{caption}
\usepackage{float}
\usepackage{dsfont}
\usepackage{hyperref}
\usepackage{xurl}

\usepackage{siunitx}
\usepackage{epsfig}

\newcommand\Tcell{\mathbf{T}_{\mathrm{cell}}}
\newcommand\Teffk{\mathbf{\tilde{T}}_{\mathrm{eff}}(\bm{k})}
\newcommand\Teff{\mathbf{T}_{\mathrm{eff}}}
\newcommand\Teffpw{\mathbf{T}_{\mathrm{eff}}^{\text{pw}}}
\newcommand\Teffdip{\mathbf{T}_{\mathrm{eff}}^\text{dip}}

\newcommand\Eq[1]{Eq.~(\ref{#1})}
\newcommand\ii{\mathrm{i}}

\begin{document}

\title{A T-matrix Based Approach to Homogenize Artificial Materials}

\author{Benedikt Zerulla}
\email{benedikt.zerulla@kit.edu}
\affiliation{Institute of Nanotechnology, Karlsruhe Institute of Technology (KIT), D-76344 Eggenstein-Leopoldshafen, Germany}
\author{Ramakrishna Venkitakrishnan}
\email{ramakrishna.venkitakrishnan@kit.edu}
\affiliation{Institute of Theoretical Solid State Physics, Karlsruhe Institute of Technology (KIT), D-76131 Karlsruhe, Germany}
\author{Dominik Beutel}
\affiliation{Institute of Theoretical Solid State Physics, Karlsruhe Institute of Technology (KIT), D-76131 Karlsruhe, Germany}
\author{Marjan Krsti\'c}
\affiliation{Institute of Theoretical Solid State Physics, Karlsruhe Institute of Technology (KIT), D-76131 Karlsruhe, Germany}
\author{Christof Holzer}
\affiliation{Institute of Theoretical Solid State Physics, Karlsruhe Institute of Technology (KIT), D-76131 Karlsruhe, Germany}
\author{Carsten Rockstuhl}
\affiliation{Institute of Nanotechnology, Karlsruhe Institute of Technology (KIT), D-76344 Eggenstein-Leopoldshafen, Germany}
\affiliation{Institute of Theoretical Solid State Physics, Karlsruhe Institute of Technology (KIT), D-76131 Karlsruhe, Germany}
\author{Ivan Fernandez-Corbaton}
\email{ivan.fernandez-corbaton@kit.edu}
\affiliation{Institute of Nanotechnology, Karlsruhe Institute of Technology (KIT), D-76344 Eggenstein-Leopoldshafen, Germany }


\begin{abstract}
The accurate and efficient computation of the electromagnetic response of objects made from artificial materials is crucial for designing photonic functionalities and interpreting experiments. Advanced fabrication techniques can nowadays produce new materials as three-dimensional lattices of scattering unit cells. Computing the response of objects of arbitrary shape made from such materials is typically computationally prohibitive unless an effective homogeneous medium approximates the discrete material. In here, we introduce a homogenization method based on the effective T-matrix, $\Teff$. Such a matrix captures the exact response of the discrete material, is determined by the T-matrix of the isolated unit cell and the material lattice vectors, and is free of spatial dispersion. The truncation of $\Teff$ to dipolar order determines the common bi-anisotropic constitutive relations. When combined with quantum-chemical and Maxwell solvers, the method allows one to compute the response of arbitrarily-shaped volumetric patchworks of structured molecular materials and metamaterials.
\end{abstract}

\maketitle

\section{Introduction and summary}
Artificial materials increase our ability to control electromagnetic fields well beyond what can be achieved with natural materials. At optical and infrared frequencies, the fabrication of deterministic photonic materials is challenging because of the small dimensions of the unit cells needed for mimicking the way nature builds materials: as periodic three-dimensional lattices containing a copy of the unit cell at each lattice point. However, recent advances in fabrication technology provide nowadays new ways to tackle exactly this challenge. In three-dimensional laser printing, the feature size resolution is reaching the sub-micrometer and nanometer scales \cite{Hahn2019}. In addition, this technique allows for the manufacturing of unit cells composed of various materials including organic (synthetic and natural) polymers, inorganic materials such as chalcogenide glasses and/or metals \cite{Yang2021}. Another promising class of artificial materials are molecular metal-organic frameworks (MOF) and their flat relatives, i.e., surface MOF (SURMOF) \cite{James03,ZHUANG2016391}. These materials feature nanometer-scale scaffold-like crystalline structures formed by organic molecules and metallic ions. A variety of geometrical shapes can be fashioned \cite{Furukawa2009,Haldar2021}, and SURMOFs with different lattices can be combined in the same functional device \cite{Oldenburg2016}.

But the benefits of technological advances can only be fully harvested if theoretical tools keep up with the pace. Regardless of the fabrication technique, the efficient and accurate simulation of the electromagnetic response of artificial materials is crucial for both the interpretation of experimental measurements and for the {\em in silico} design of new materials and devices. In this context, a particularly useful formalism is the T-matrix or transition matrix formalism \cite{Waterman1965,Mishchenko2020}, which, for linear light-matter interactions, produces the field scattered off a given object under general illumination. When dealing with an infinite periodic repetition of a unit cell, the calculation of the lattice couplings is very conveniently done \cite{eyert2012} using the Ewald summation method \cite{Ewald21}. The T-matrix and Ewald's method can be combined in numerical codes for computing the electromagnetic response of infinitely periodic systems \cite{stefanou1998,stefanou2000,Beutel:21}, achieving efficiencies more than two orders of magnitude better than numerical solvers of Maxwell differential equations \cite{jin2015finite,taflove2005computational}. The calculations of the T-matrices of molecular unit cells by quantum-mechanical {\em ab initio} methods \cite{Fernandez-Corbaton:2020}, in particular time-dependent density-functional theory (TD-DFT) \cite{SURMOFCavity,doi:10.1021/ja991960s}, enable the consideration of systems including slabs of molecular materials such as optical planar cavities filled with SURMOFs \cite{SURMOFCavity}. Unfortunately, Ewald's method cannot be used for finite arrangements of scatterers, such as an object of finite shape made from a 3D lattice of unit cells. Many of these objects cannot be handled by other T-matrix based methods either, because of the computational cost when the number of unit cells grows beyond a few thousands. This then excludes, for example, non-planar arrangements of meta-atoms \cite{Ergin2010}, and finite objects made from molecular materials \cite{Furukawa2009,Haldar2021}. Additionally, the co-existence of slabs with different lattice vectors in the same system \cite{Oldenburg2016} is also an obstacle for Ewald's summations.

For many years, research based on effective medium theories is aiming at alleviating the computational burden by replacing the discrete lattice of scatterers by a homogeneous effective medium \cite{sihvola2002electromagnetic,agranovich2013crystal,smith2006homogenization,silveirinha2006nonlocal,hossain2022homogenization,alu2011first,Sihvola92,PhysRevB.75.115104,pendry2000negative,Monti:11,rockstuhl2013amorphous,JEOS:RP06019,PhysRevB.77.233104,photonics2020540,soukoulis2007negative,Yan:09,Ishimaru2003,PhysRevB.91.155406,Papadakis18,maslovski2002wire,mnasri2019homogenization,tretyakov2016personal,PhysRevE.75.036603,chebykin2015spatial,doi:10.1063/1.1305828,belov2005homogenization,PhysRevB.76.245117,liu2011limitations,ciattoni2015nonlocal,silveirinha2007metamaterial,repan2021artificial}. Homogenization theories address the computation of the material parameters of such an effective medium for a given constitutive relation. Then, ideally, the electromagnetic response of a target object of arbitrary shape made out of the actual discrete material should be well approximated using the constitutive relations of the effective medium, plus appropriate boundary conditions. In this way, much larger finite-size objects can be considered compared to what is possible without homogenization.

Homogenization is a complex endeavor, and its complexity is reflected as different kinds of shortcomings in different homogenization techniques. Simpler approaches can even have contradicting conditions for their applicability: For example, they often require lattice constants much smaller than the wavelength and a sufficiently sparse lattice simultaneously. Also, long wavelength approximations in the calculation of the response of the unit cells and/or the lattice couplings are often made. A salient problem of some advanced methods is spatial dispersion: the dependence of the set of material parameters on the direction of the wavevector of the illumination. This is an important limitation since it is then unclear how to use the many-fold instances of the effective material parameters in practice, except possibly for planar geometries where only a few propagation directions are involved. Some homogenization methods are based on retrieval, where a reference object made with the actual material, typically a slab, is probed with different illuminations. The numerically measured response is then fitted, e.g., in a least-square sense, by the predictions of a homogeneous model of the reference object whose material parameters are varied in an optimization procedure. While, by design, retrieval methods do not suffer from spatial dispersion, a potentially weak point is that a reference object is involved from the beginning, and it is not immediately obvious that the retrieved material parameters can be used for target objects with different shapes. For example, machine learning techniques confirm a non-uniqueness issue \cite{simovski2010electromagnetic} showing that, in some cases, the measurements can be well approximated by different sets of material parameters \cite{repan2021artificial}. Whether all the different sets are valid for different objects remains unclear. We are not aware of any homogenization technique that is free of all the aforementioned shortcomings. Moreover, most techniques also lack an {\em a priori} quality control mechanism independent of shape. Without such mechanism, assessing the accuracy of the homogeneous model for a particular target object would involve the comparison with simulations that explicitly consider the discrete lattice, hence defeating the purpose of homogenization.

In this article, we introduce a novel homogenization method whose starting point is the non-spatially dispersive yet exact response of the material, and where the material parameters of the constitutive relations are determined from the dipolar part of the response without considering any particular shape of a target object. The quantification of the difference between the exact description and its dipolar part constitutes a built-in quality metric that {\em a priori} indicates the suitability or unsuitability of using the homogeneous model.

The central object of our novel homogenization method is a linear operator that provides an exact description of the linear interaction of light with the bulk material, that is, with the infinite 3D lattice of scatterers. The linear operator has the form of a T-matrix in the multipolar basis, which we call the effective T-matrix: $\Teff$. The effective T-matrix is computed using the lattice vectors  to obtain the mutual interaction and the T-matrix of a single isolated copy of the unit cell, which we will call $\Tcell$. All the couplings due to the infinite lattice are incorporated in $\Teff$. Such couplings change $\Tcell$ into $\Teff$ while, at the same time, removing the lattice interactions. In other words, one can equivalently describe the response of the material by replacing the copies of $\Tcell$ interacting with each other with copies of $\Teff$ which do not interact with each other, i.e. they are invisible to each other. The effective T-matrix is an excellent starting object for homogenization because it is independent of any target object shape, it decouples the unit cells, and is an exact description of the interaction of light with the 3D lattice of scatterers of the actual material. Importantly, $\Teff$ does not suffer from spatial dispersion.

We show that the dipolar part of $\Teff$, which we call $\Teffdip$ and has 36 parameters, is bijectively connected with a very common 6$\times$6 model for the constitutive relations of the effective medium, which is complemented by the usual (bi-anisotropic) boundary conditions \cite[Section~4.3]{Serdiukov2001}. The contributions of higher multipolar orders contained in $\Teff$ are excluded from this homogeneous model. This is the only point where our model deviates from the exact response. The effective material parameters in the given constitutive relations derived in this way: i) are completely determined by the kind of 3D lattice and scatterers in the unit cell without any influence from the shape of any target object, ii) contain all the modifications that the lattice causes to the dipoles, and iii) do not suffer from spatial dispersion. The material parameters can be used in software packages such as COMSOL Multiphysics \cite{Comsol}. A target object made as a volumetric patchwork of domains with different discrete materials can also be considered. When compared to other methods, neither retrieval nor fitting procedures are needed, and the actual assignment of effective properties is a straightforward computation using the T-matrix framework.  Within one calculation, all entries of the effective material tensors are computed.

Even before calculating $\Teff$, the band diagram of the actual material is used to judge whether the material can be homogenized at all. For example, the homogeneous model is clearly inadequate for frequencies that produce diffraction in the lattice, as X-rays produce in most solids. But even before this obvious limit, light starts to probe the presence of the lattice due to Bragg reflections.

We show that the response of slabs and spheres made of the actual discrete material are very accurately approximated by this method if the material can be homogenized at all and the relative matrix distance between $\Teff$ and $\Teffdip$ is small. 

The rest of the paper is organized as follows. The proposed homogenization method is explained in Section~2, and Sections~3 to 5 contain different application examples. In Section~3, the material is a cubic lattice of gold spheres that is homogenizable in the considered frequency range, and whose $\Teff$ and $\Teffdip$ are essentially identical. The results of the homogeneous model in a slab and a sphere match very well the corresponding exact solutions that explicitly consider the discrete lattice. In Section~4, the material is a cubic lattice of cut-plate pairs, which is only homogenizable in the lower part of the considered frequency range, and for which the difference between $\Teff$ and $\Teffdip$ is two orders of magnitude larger than in the previous material, and exhibits a growing trend with the frequency. In this case, the homogeneous model predicts the exact results reliably only in the lower part of the considered range of frequencies. Section~5 contains the application to a chiral SURMOF which features anisotropic chirality. Homogenization produces essentially a perfect match with the exact results for a slab of the SURMOF. Then, the circular dichroism (CD) of an array of spheres made from the molecular material is computed in COMSOL for perpendicular and oblique illumination directions. The CD is much larger at oblique incidence. This last prediction is possible and trustworthy only because of the accurate homogeneous model. Section~6 contains the conclusion and outlook.

\section{Homogenization based on the effective T-matrix of a material}

\begin{figure*}[t!]
\centering
	\includegraphics[width=0.95\textwidth]{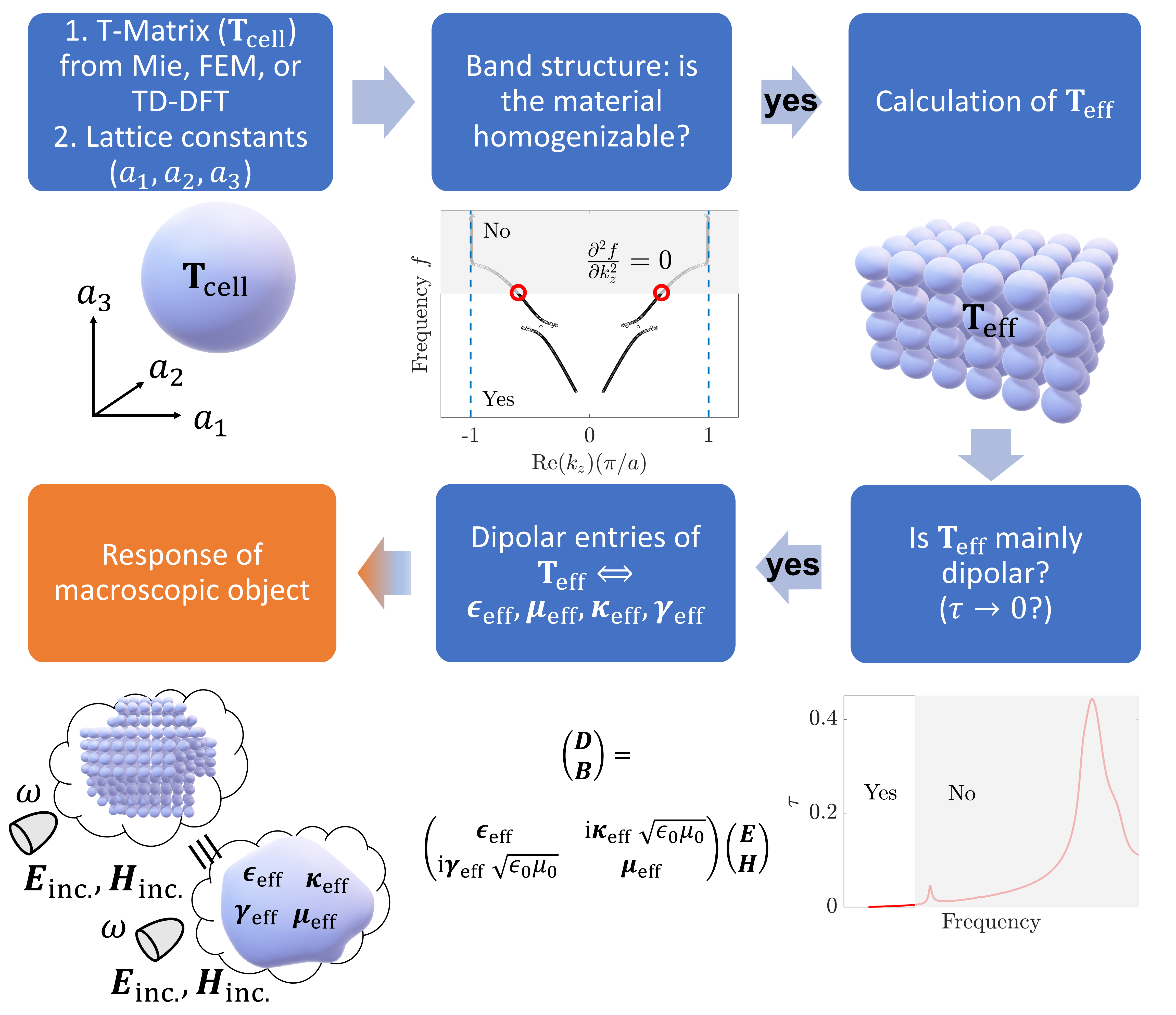}
	\caption{The blue blocks describe our novel homogenization method that is completely independent of the shape of the final target object in the orange block. The starting point is the upper left corner, where we calculate the T-matrix of the unit cell from which the material is made. The unit cell can contain molecules and macroscopic scatterers, whose T-matrices are computed with TD-DFT and Maxwell solvers, respectively. Afterward, there is a straight flow along the arrows to advance. The blocks with question marks represent two criteria that must be met to ensure accurate results: that the light does not feel the discrete lattice, and that, after accounting for all the lattice interactions, the response of a unit cell ($\Teff$) is mostly dipolar. The dipolar part of $\Teff$ is bijectively connected to the bi-anisotropic 6$\times$6 local constitutive relations model for a homogeneous medium.}
    \label{fig:Workflow}	
	\end{figure*}
Figure \ref{fig:Workflow} is a block diagram of our homogenization method. The discrete bulk material is defined by an infinite 3D lattice and a unit cell, repeated at each lattice point. The lattice is defined by its three lattice vectors. The unit cell is defined by $\Tcell$, the T-matrix of the scatterer(s) composing the unit cell. This T-matrix relates the electromagnetic fields incident upon an isolated unit cell outside the lattice to the corresponding scattered electromagnetic fields. The T-matrix formalism was introduced by Waterman \cite{Waterman1965} and is nowadays a popular tool in physics and engineering \cite{Mishchenko2020}. The T-matrix of an isolated scatterer can be calculated by methods such as Mie theory for spheres, or the finite-element method (FEM) \cite{Fruhnert2017,PhysRevB.99.045406} or the Extended Boundary Conditions Method (EBCM) \cite{NIEMINEN20031019} for more complicated objects. Most commonly, the multipolar basis of vector spherical waves is used to expand the incident and scattered fields. The size of the T-matrix becomes finite by truncating such expansions to some maximum multipolar order while ensuring that the contribution of the discarded higher orders to the light-matter interaction is negligible.  For individual molecules or molecular clusters, the T-matrix can be obtained using {\em ab initio} quantum chemical methods \cite{Fernandez-Corbaton:2020} such as TD-DFT. The T-matrix unifies the description of light-matter interactions for both molecules and macroscopic objects.

After the definition of the actual material, the first question to answer is whether the material can be homogenized at all at the frequencies of interest. This question is independent of the specific homogenization approach. For example, the homogeneity assumption clearly breaks down at frequencies that produce diffraction in the lattice, akin to X-rays in most solids. But even before this obvious limit, the material can act as a photonic crystal due to Bragg reflections in the lattice. For a cubic lattice with lattice constant $a$, diffraction starts at a propagation constant $\beta = 2\pi /a$, while the first Bragg reflection occurs already at the edge of the Brillouin zone at a propagation constant of $\beta = \pi /a$. We note that $\beta$ is the propagation constant of some fundamental (Bloch) mode propagating in the periodically structured material.

Therefore, as in \cite{Ryb15}, we use the band structure of the actual material to determine whether the material can be homogenized. When the wavenumber obtained from the band structure approaches the edge of the Brillouin zone, the onset of a Bragg band gap can be clearly seen [e.g., in Figure~\ref{fig:HomCPP}]. The presence of the band gap will start affecting the response of the material already at smaller frequencies by bending the dispersion relation. Starting around the point that the second derivative of the dispersion relation vanishes, the light is explicitly affected by the lattice, and the results of homogenized models will hence become increasingly unreliable. The question of whether homogenization is feasible or not can, therefore, be judged by inspecting the emerging band structure. The band structure can be calculated by solving the eigenvalue equation of the material with a full-wave solver such as mpGMM once the T-matrix of the object is known and the lattice geometry is fixed \cite{Beutel:21}.

\begin{figure*}[t!]
\centering
   
	\includegraphics[width=0.8\textwidth]{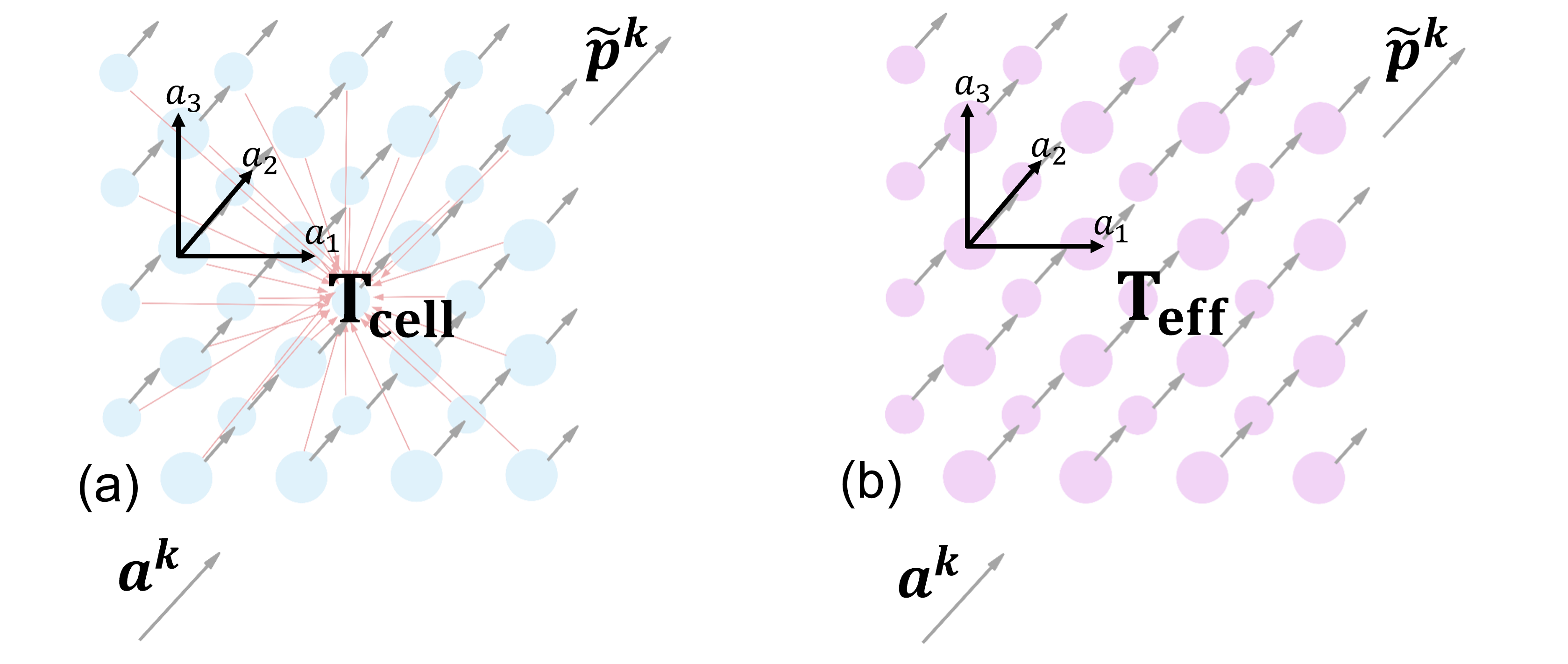}

\caption{Two equivalent descriptions of light-matter interaction in a 3D lattice of identical scatterers, where an excitation $\bm{a^k}$ produces a response $\bm{\tilde{p}^k}$. \textbf{(a)} Interacting scatterers in a lattice represented by their individual T-matrices $\Tcell$. \textbf{(b)} Scatterers in a lattice represented by their effective T-matrix $\Teff$. Every scatterer has the same $\Teff$, which incorporates all the lattice interactions. As a result, the scatterers represented by $\Teff$ are invisible to each other. The effective T-matrix is an excellent starting point for homogenization.}
    \label{fig:TeffIllustration}
	\end{figure*}

When homogenization is possible, the response of the actual material, $\Teff$, is computed from the 3D lattice and the T-matrix of the unit cell, $\Tcell$. One salient feature of $\Teff$ is that it is an exact description of the material response as long as enough multipolar orders are considered in its calculation. Another salient feature is that $\Teff$ does not suffer from spatial dispersion. The effect of the lattice is different for different illumination directions, but it is possible to rigorously obtain a single object, $\Teff$, valid for all directions. Before going into the details of the calculation, it is beneficial to understand the physical meaning of $\Teff$, illustrated in Figure~\ref{fig:TeffIllustration}: A 3D lattice of scatterers described by $\Tcell$, which interact among each other in Figure~\ref{fig:TeffIllustration}(a), is rendered equivalent to the same lattice but with a different unit cell described by $\Teff$ in Figure~\ref{fig:TeffIllustration}(b). The new ``objects'' are invisible to each other because all the lattice interactions have been included in $\Teff$. Therefore $\Teff$ collects all non-local effects into an effectively local description where, as seen in Figure~\ref{fig:TeffIllustration}, the response of a unit cell is independent of all other unit cells. Let us now examine the details.

We start by considering Equation~(17) of Ref.~ \onlinecite{Beutel:21}, which is the expression for an effective T-matrix $\mathbf{\tilde{T}}_{\mathrm{eff}}(\bm{k}_{\parallel})$ describing the scattering by an object located at the origin of a 2D periodic lattice, and including all the lattice couplings. Such an effective T-matrix depends on the propagation direction of the incident light through the component of the wave vector parallel to the lattice plane, $\bm{k}_{\parallel}$. For a 3D lattice, the formula is identical except that $\bm{k}_{\parallel}$ is replaced by the total wave vector $\bm{k}$ \cite{Beutel2022}:

\begin{equation}\label{eq:Teff}
    \mathbf{\tilde{T}}_{\mathrm{eff}}(\bm{k})=\left(\mathbf{I}-\Tcell\sum_{\bm{R}\neq 0}\mathbf{C}^{(3)}(-\bm{R})\mathrm{e}^{\mathrm{i}\bm{k}\cdot\bm{R}}\right)^{-1}\Tcell,
\end{equation}

\noindent where $\Tcell$ is the T-matrix of an isolated unit cell of the lattice. The matrix $\mathbf{\tilde{T}}_{\mathrm{eff}}(\bm{k})$ connects $\bm{a}(\bm{k})$, the multipolar expansion coefficients of the original plane wave incident on the unit cell located at the origin to $\bm{\tilde{p}}$, the multipolar expansion coefficients of the corresponding scattered field

\begin{align}\label{eq:ScatCoeff}
\bm{\tilde{p}}= \Teffk\bm{a}(\bm{k}).   
\end{align}
The definition of the multipolar expansion functions, also known as multipolar fields or vector spherical harmonics, can be found in Equations~(15, S3a-S3d) of Ref.~\onlinecite{Beutel:21}. The $\mathbf{C}^{(3)}(-\bm{R})$ matrices in \Eq{eq:Teff} represent the electromagnetic coupling between the origin and the $\bm{R}$ lattice point, and their elements are the translation coefficients for vector spherical waves [see e.g. Equations~(S6a-S7) of Ref.~\onlinecite{Beutel:21}]. The infinite sum $\sum_{\bm{R}\neq 0}\mathbf{C}^{(3)}(-\bm{R})\mathrm{e}^{\mathrm{i}\bm{k}\cdot\bm{R}}$ over all the lattice points except for the origin is computed with Ewald's summation method \cite{Beutel:21,Ewald21,Kambe67}. The sum represents the total electromagnetic coupling between the unit cell in $\bm{R}=\mathbf{0}$ and all the other unit cells in the infinite 3D lattice. The computation of the coupling is exact up to the fact that a maximum allowed multipolar order must be selected. We note that, in contrast to approaches that use the quasi-static approximation, see \cite{doi:10.1063/1.1305828}, for instance, the coefficients in $\mathbf{C}^{(3)}(-\bm{R})$ are exact, depend on the wavenumber $k$, and take the spatial oscillations of the fields into account. 

The $\Teffk$ matrix in \Eq{eq:ScatCoeff} has an important disadvantage regarding its use as the starting point for homogenization: It is only appropriate for a particular field, namely a plane wave with momentum $\bm{k}$. Therefore, $\Teffk$ should not be used for any other incident direction. This can be appreciated from the fact that the 3D lattice ``looks differently when looked at from different directions'', which impacts the lattice sums through the $\mathrm{e}^{\mathrm{i}\bm{k}\cdot\bm{R}}$ factor. If $\mathbf{\tilde{T}}_{\mathrm{eff}}(\bm{k})$ is used to derive material parameters, those would depend on the directions of $\bm{k}$. Such dependence is sometimes called {\em spatial dispersion}. It is then unclear how to use the many-fold instances of material parameters, except possibly for planar slabs where only a few plane wave directions are involved. Fortunately, this problem can be solved rigorously.

The physical ideas behind our solution to such problem can be stated as follows. Let us assume that we decompose the scattered field $\bm{\tilde{p}}$ in \Eq{eq:ScatCoeff} into plane waves. Then, \Eq{eq:ScatCoeff} can be seen as providing one of the columns of a T-matrix in the plane wave basis $\Teffpw$: The system is excited by a plane wave and produces scattered plane waves. Now, the entire $\Teffpw$ can be obtained by scanning the direction of $\bm{k}$. Once this is done, $\Teffpw$ can be changed from the plane wave basis to the multipolar basis to obtain $\Teff$, an effective T-matrix in the multipolar basis that is (i) valid for all $\bm{k}$ directions and (ii) not explicitly dependent on the $\bm{k}$ direction. The Methods section contains an analytical derivation that formalizes these ideas into formulas for the computation of $\Teff$. Besides that analytical approach, the matrix $\Teff$ can also be computed by adapting the procedure introduced in Ref.~ \onlinecite{Fruhnert2017}. First, a finite number of points on the $\bm{\hat{k}}$ sphere, i.e., on the sphere of directions of $\bm{k}$, is selected. A particularly useful method for selecting equally-spaced points on a sphere can be found in \cite{partSamp}. Then, the $\Teffk$ matrices corresponding to each $\bm{\hat{k}}$ are computed, and \Eq{eq:ScatCoeff} is used two times for each $\bm{\hat{k}}$, one for each of the two possible polarization handedness of an incident plane wave with momentum $\bm{k}$. Then, the coefficients of all the incident plane waves $\bm{a}^{\bm{k}_1},\dots,\bm{a}^{\bm{k}_K}$, and their corresponding expansions of the scattered waves $\bm{\tilde{p}}^{\bm{k}_1},\dots,\bm{\tilde{p}}^{\bm{k}_K}$, are collected in the following matrix equation

\begin{align}\label{eq:ModTeff}
    (\bm{\tilde{p}}^{\bm{k}_1},\dots,\bm{\tilde{p}}^{\bm{k}_K})=\mathbf{T}_{\mathrm{eff}}(\bm{a}^{\bm{k}_1},\dots,\bm{a}^{\bm{k}_K}),
\end{align}
from where $\Teff$ can be obtained. The number of points in the directions sphere should be much larger than the size of $\Teff$. The key aspect of \Eq{eq:ModTeff} is that it imposes that $\Teff$ shall respond to an incident plane wave with a specific propagation direction $\bm{\hat{k}}$ as $\mathbf{\tilde{T}}_{\mathrm{eff}}(\bm{k})$ responds. As Figure~S3 shows, the $\Teff$ obtained from \Eq{eq:ModTeff}, which we use in the examples contained in this article, is essentially identical to the $\Teff$ obtained from the direct implementation of the analytical formulas in the Methods section.

With $\Teff$ at hand, the next step is to choose a homogeneous model for the effective medium and then use $\Teff$ for extracting the parameters of the model. In this work, we choose the linear 6$\times$6 local bi-anisotropic model, where the constitutive relations connecting the electric displacement $\bm{D}$ and magnetic flux density $\bm{B}$ to the $\bm{E}$ and $\bm{H}$ fields in the effective homogeneous medium read in frequency domain [see e.g., Eq~(1.51) in Ref.~\onlinecite{Kristensson2016}]

\begin{align}\label{eq:ConsRel}
	\begin{pmatrix}
		\bm{D}(\omega)\\
		\bm{B}(\omega)
	\end{pmatrix}
	=
	\begin{pmatrix}
	\bm{\varepsilon}_{\mathrm{eff}}(\omega)&\mathrm{i}\bm{\kappa}_{\mathrm{eff}}(\omega)\sqrt{\varepsilon_0\mu_0}\\
	\mathrm{i}\bm{\gamma}_{\mathrm{eff}}(\omega)\sqrt{\varepsilon_0\mu_0}&\bm{\mu}_{\mathrm{eff}}(\omega)
	\end{pmatrix}
	\begin{pmatrix}
	\bm{E}(\omega)\\
	\bm{H}(\omega)
	\end{pmatrix},
	\end{align} 
where $\bm{\varepsilon}_{\mathrm{eff}}(\omega)$ is the tensorial permittivity, $\bm{\mu}_{\mathrm{eff}}(\omega)$ the tensorial permeability, and the $\bm{\kappa}_{\mathrm{eff}}(\omega)$ and $\bm{\gamma}_{\mathrm{eff}}(\omega)$ tensors describe the coupling between the electric and magnetic fields. Such constitutive relations are the most general local and linear constitutive relations. Reciprocal materials such as the ones that we will consider in this article meet $\bm{\gamma}_{\mathrm{eff}}=-\bm{\kappa}^{\mathrm{T}}_{\mathrm{eff}}$ \cite{Kristensson2016}. 

The model in \Eq{eq:ConsRel} with its corresponding boundary conditions is very commonly used and has full or partial support in popular Maxwell solvers. For example, it can be implemented in COMSOL Multiphysics \cite{Comsol}, which supports the full 6$\times$6 model. In MEEP \cite{MEEP}, the anisotropic magneto-electric couplings are restricted to have a particular structure, and CST \cite{CST} supports anisotropy in the electric-electric and magnetic-magnetic tensors but not in the magneto-electric ones. 

The Methods section contains the derivation that bijectively connects the 6$\times$6 effective constitutive matrix in \Eq{eq:ConsRel}, to {\em the dipolar part of $\Teff$}, which we denote by $\Teffdip$ and also has 36 parameters. Formally, $\Teffdip$ can be seen as the result of zeroing out all the entries of $\Teff$ except those relating incident dipolar fields with scattered dipolar fields. The connection reads: 

\begin{align}\label{eq:EffParamIntro}
\begin{pmatrix}
\bm{\varepsilon}_{\mathrm{eff}}&\mathrm{i}\bm{\kappa}_{\mathrm{eff}}\sqrt{\varepsilon_0\mu_0}\\
\mathrm{i}\bm{\gamma}_{\mathrm{eff}}\sqrt{\varepsilon_0\mu_0}&\bm{\mu}_{\mathrm{eff}}
\end{pmatrix}
&=
\begin{pmatrix}
{\varepsilon}_{\mathrm{h}}\mathbf{I}_3&0\\ \nonumber
0&{\mu}_{0}\mathbf{I}_3
\end{pmatrix}+\\ 
&+n\left(\mathbf{I}_6-n\cdot q\ s_1\left[\Teffdip,\mathbf{L}\right]\right)^{-1}\times \\
&\times
	q\ s_2\left[\Teffdip\right]\mathrm{,}\nonumber
\end{align}
where $\mathbf{L}$ is a depolarization tensor, and we have dropped the explicit $\omega$-dependence for the benefit of a more concise notation. See \Eq{eq:EffParam} in Methods for the definitions of the elements in \Eq{eq:EffParamIntro}, including the functions $s_1\left[\cdot,\cdot\right]$, and $s_2\left[\cdot\right]$.

The term that contains $\mathbf{L}$ represents the depolarization of a lattice of non-interactive scatterers. This is different in other methods \cite{Sihvola92,Ishimaru2003}, where the depolarization is due to the interaction between the scatterers.

It is important to note that $\Teffdip$ contains contributions from the dipolar {\em and the non-dipolar} parts of the T-matrix of the isolated scatterer $\Tcell$. The latter contributions originate from multipolar couplings in the lattice and can be very significant in dense lattices even for electromagnetically small objects [see Figure~\ref{fig:GoldSpheresR1nmA205nm}(c)].

The local non-spatially dispersive material parameters contain all the modifications that the non-local lattice interactions produce to the dipolar response. The frequency-dependent formulation accommodates any existing temporal dispersion. 

Crucially, the bijective connection in \Eq{eq:EffParamIntro} provides a criterium to check the suitability of the homogeneous material model assumed in \Eq{eq:ConsRel}: The non-dipolar terms in $\Teff$ must be negligible. This can be quantified with the following formula:

\begin{equation}
	\label{eq:diptest}
	\tau\left(\Teff\right)=\sqrt{\frac{\mathrm{Tr}\left\{\left(\Teffdip-\Teff\right)^{\dagger}\left(\Teffdip-\Teff\right)\right\}}{2\left(\mathrm{Tr}\left\{{\Teffdip}^\dagger\Teffdip\right\}+\mathrm{Tr}\left\{\Teff ^{\dagger}\Teff\right\}\right)}},
\end{equation}

where $\tau\left(\Teff\right)\in [0,1]$, $\tau\left(\Teff\right)=0$ implies $\Teff=\Teffdip$, and $\dagger$ denotes transpose conjugate. To calculate $\tau\left(\Teff\right)$, $\Teffdip$ consists of the effective dipolar part in the upper-left corner and otherwise of entries equal to zero so that the dimensions of $\Teffdip$ and $\Teff$ are the same. We note that $\sqrt{\text{Tr}\left\{{\mathbf{A}}^\dagger \mathbf{A}\right\}}$ is the square root of the sum of the squared absolute value of each individual entry of the matrix $\mathbf{A}$, which is the expression of the Hilbert-Schmidt norm of $\textbf{A}$. A very small $\tau\left(\Teff\right)\rightarrow0$ is needed to ensure the suitability of \Eq{eq:ConsRel}. 

While the second block in Figure~\ref{fig:Workflow} determines whether the actual material is at all homogenizable, $\tau\left(\Teff\right)\rightarrow 0$ indicates that the particular model in \Eq{eq:ConsRel} is sufficient for obtaining accurate results in the end.

We emphasize that the shape of a target object never enters the computation of $\Teff$, $\Teffdip$, or the computation of the effective material parameters in the constitutive relations. This is in sharp contrast to retrieval homogenization approaches, where the effective material parameters are obtained by fitting the response of a reference object made from the actual material. Here, only the bulk material is considered, and all properties are derived from it.

The methodology and its limits are illustrated in the next sections with slabs,  spheres, and arrays of spheres made from different materials. In particular, the applicability to molecular materials is demonstrated for a SURMOF featuring anisotropic chirality. The examples show the value of the two homogenization criteria, which determine whether a particular material is homogenizable with \Eq{eq:EffParam} at a particular frequency: When the two homogenization criteria are simultaneously satisfied, the electromagnetic response of a target object made from the actual material can, independently of the shape of the object, be computed very precisely using the constitutive relations in \Eq{eq:EffParam}. Crucially, both criteria can be tested {\em before} any simulation of the target object.

\section{Gold spheres in a cubic lattice}
 As a first example, we consider gold spheres of \SI{1}{\nano\meter} radius arranged in a cubic lattice with lattice constant $a=\SI{2.05}{\nano\meter}$. The surrounding host medium has a relative permittivity of ${\varepsilon}_{\mathrm{r,h}}=2.25$ \cite{ROC08}. The material parameters of gold are taken from \cite{PhysRevB.6.4370}. All multipoles up to the $N=5$ multipolar order are included in the calculations. The material is chosen as an example where homogenization is certainly feasible. First, the T-matrix of a single
 \begin{figure*}
\centering
\subfloat{
	\includegraphics[width=0.45\textwidth]{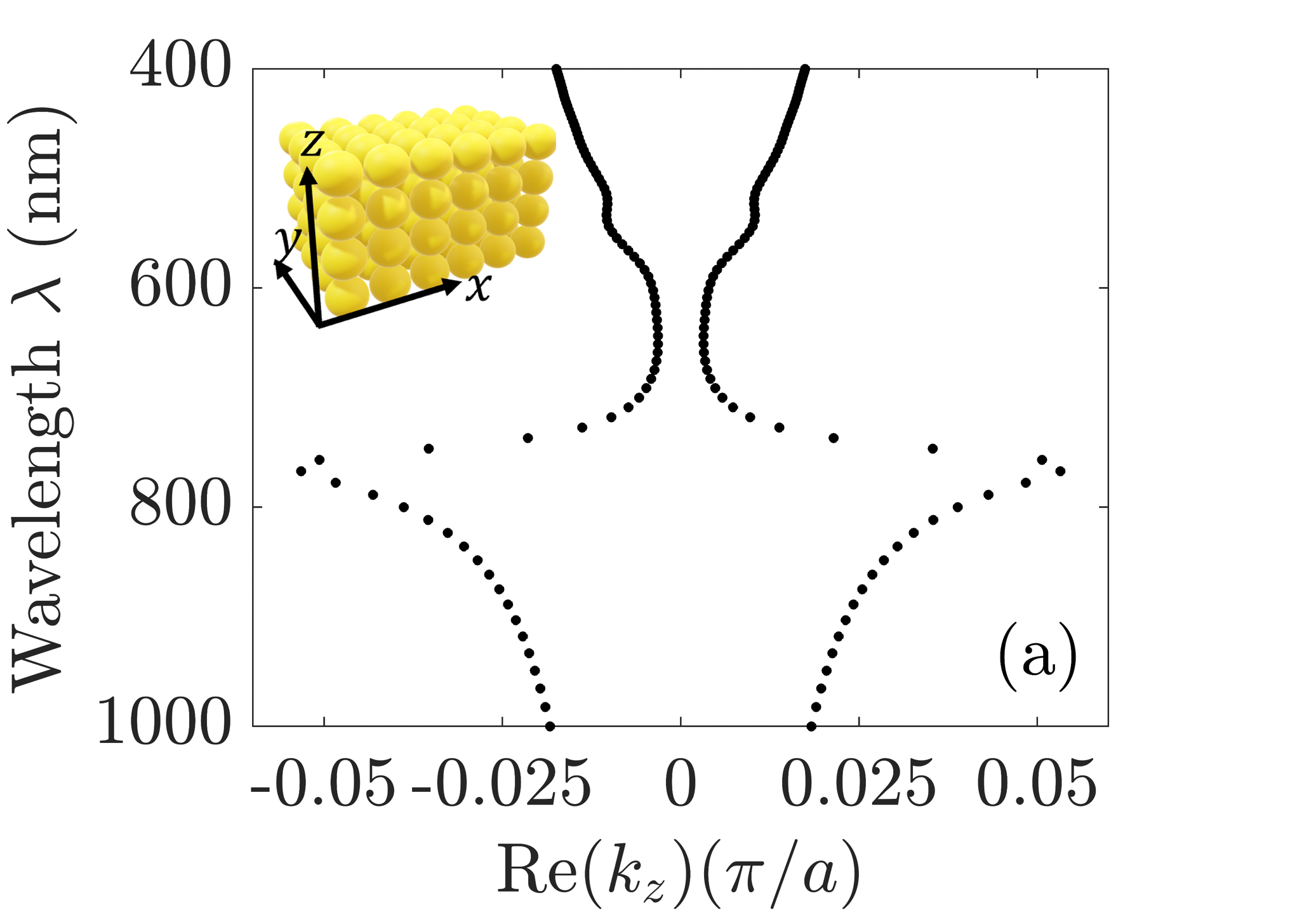}
	}
	\subfloat{
	\includegraphics[width=0.45\textwidth]{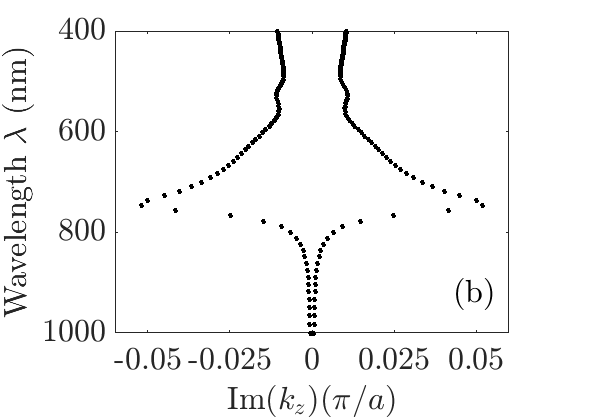}
	}\\
   \subfloat{
	\includegraphics[width=0.45\textwidth]{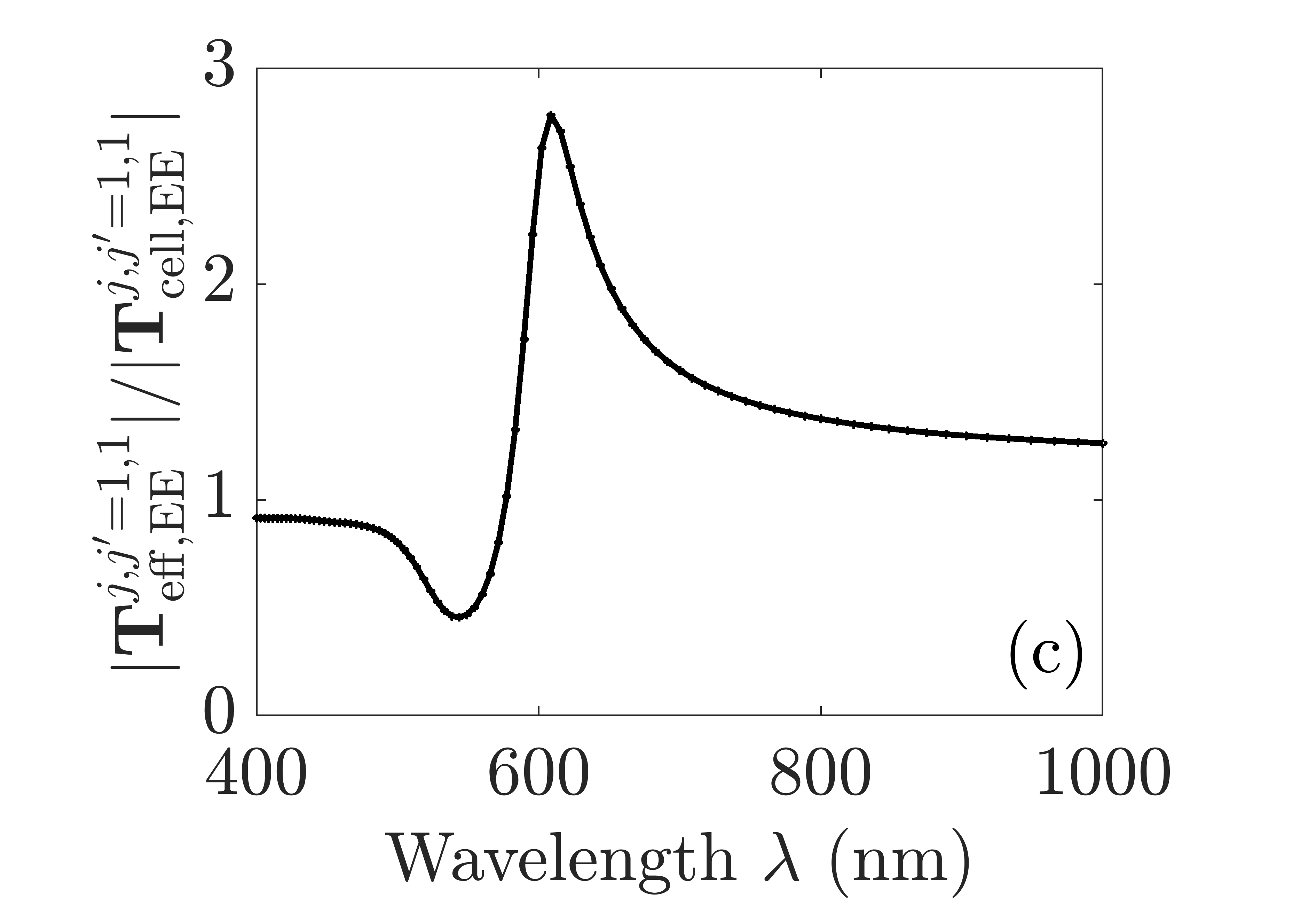}
	}
	\subfloat{
	\includegraphics[width=0.45\textwidth]{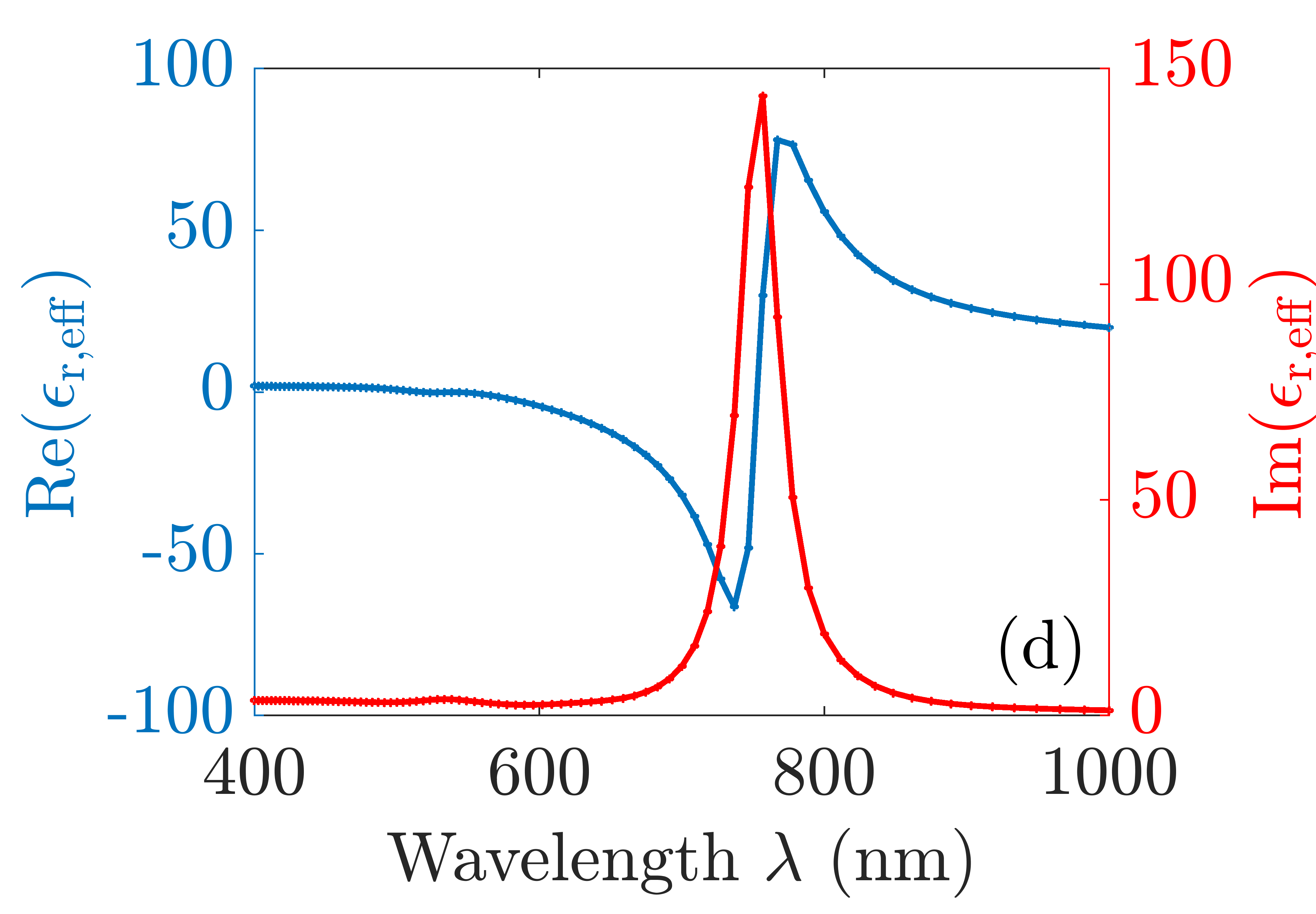}
	}\\
	\subfloat{
	\includegraphics[width=0.45\textwidth]{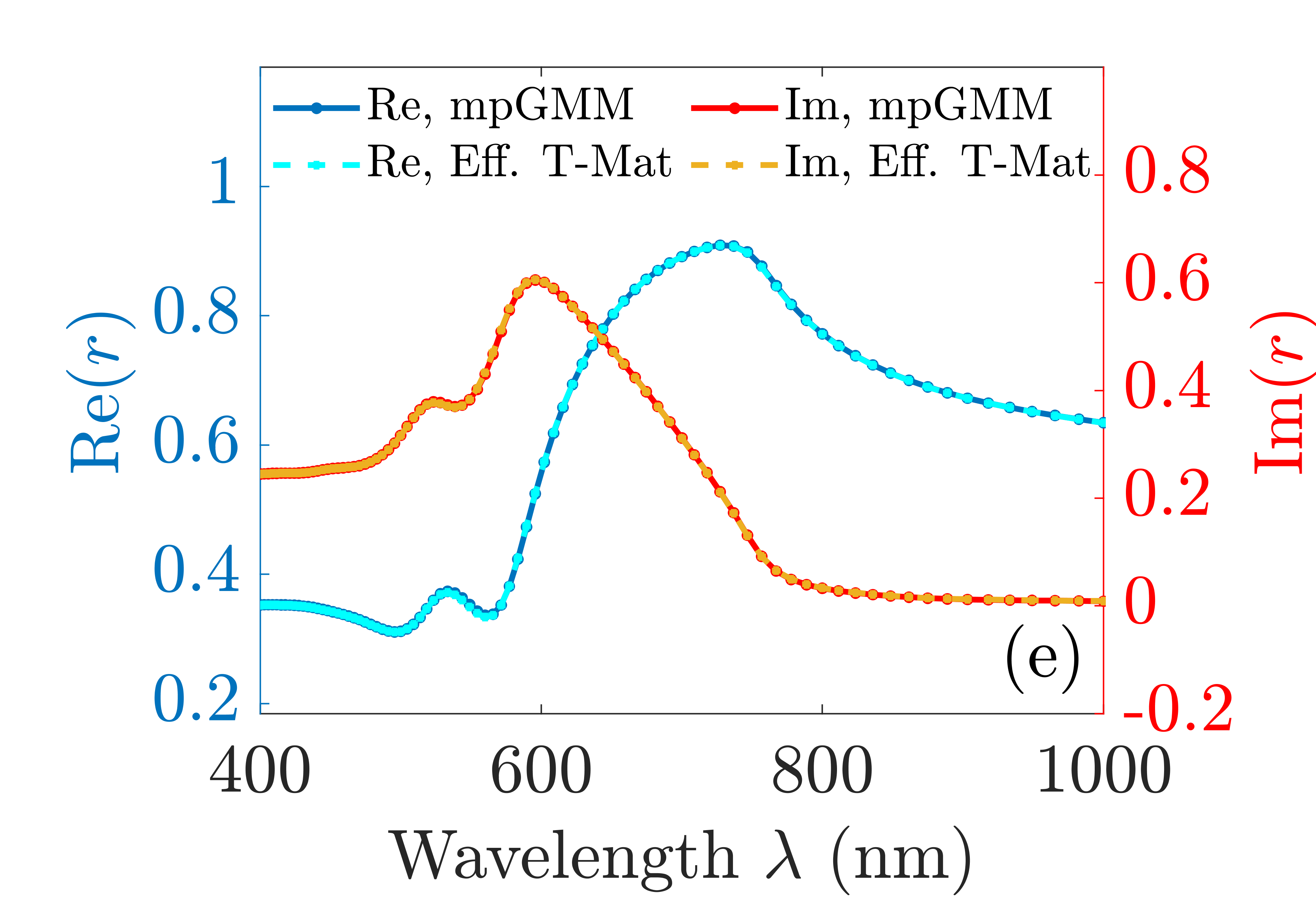}
	}
	\subfloat{
	\includegraphics[width=0.45\textwidth]{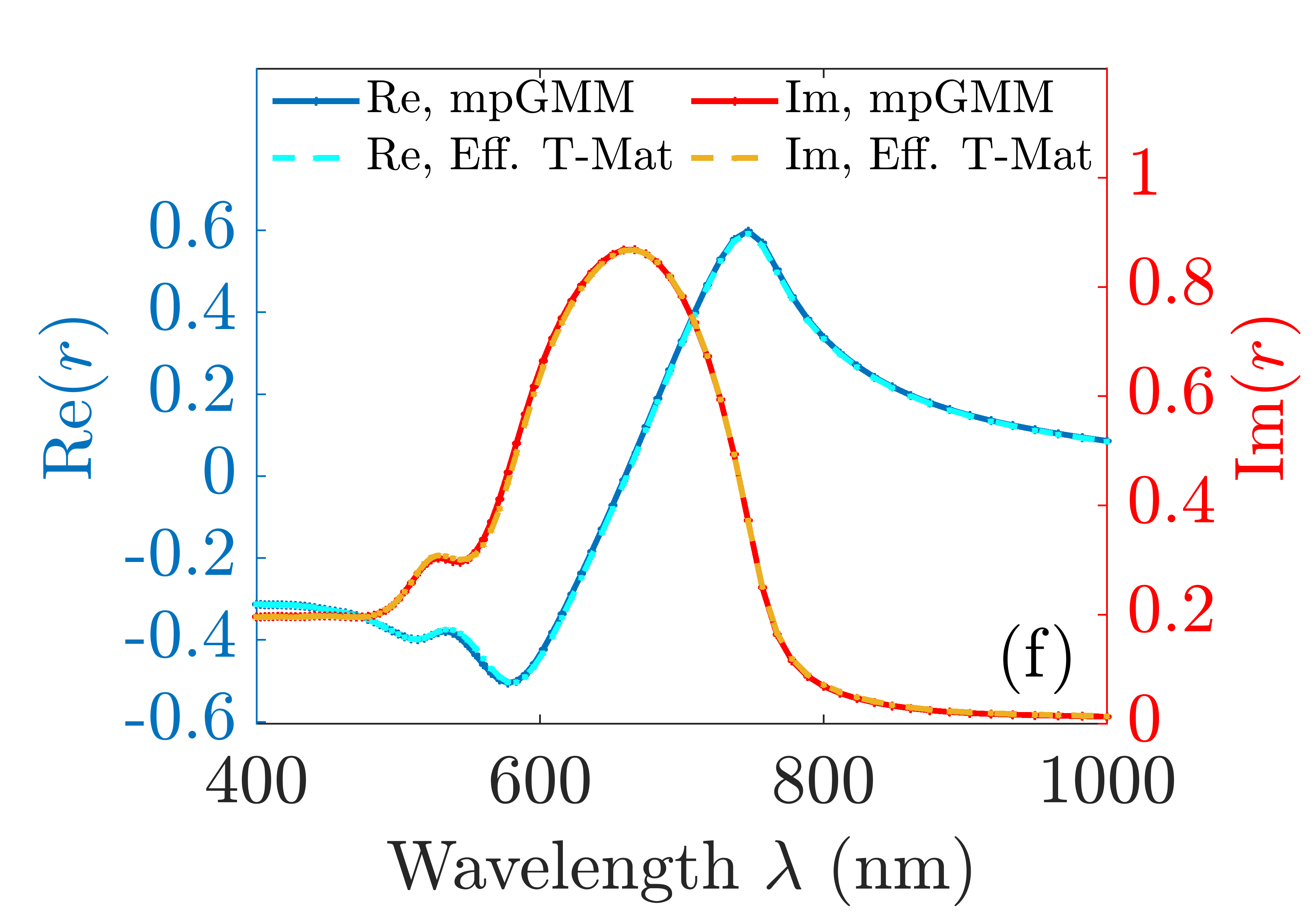}
	}
	\caption{Band structure for real \textbf{(a)}  and imaginary \textbf{(b)} parts $k_z$ of a 3D lattice of gold spheres (inset) with a radius of \SI{1}{\nano\meter} arranged in a cubic lattice with a lattice constant of $a=\SI{2.05}{\nano\meter}$. The bands are bent due to an electric dipolar resonance far away from the edge of the Brillouin zone so that the material can be homogenized. \textbf{(c)} Ratio between the electric dipolar entry of the $\Teff$ of the material and the same T-matrix entry of the isolated sphere. \textbf{(d)} Effective permittivity of the lattice structure obtained with Equation~(\ref{eq:EffParamIntro}). Transverse Magnetic (TM) reflection coefficient for normal incidence \textbf{(e)} and for an oblique incidence \textbf{(f)} of a \SI{2.15}{\milli\meter} thick slab, corresponding to $2^{20}$ layers of gold spheres.  For oblique incidence, the direction of the wave vector is $\bm{\hat{k}}_{\mathrm{inc}}=[\sin(\theta),0,\cos(\theta)]^{\mathrm{T}}$ with $\theta=75^{\circ}$. We observe in both cases that there is a perfect agreement between the results obtained with the effective parameters and the exact results for the actual non-homogeneous slab made from the lattice of gold spheres calculated with the full-waver solver mpGMM.}
    \label{fig:GoldSpheresR1nmA205nm}	
	\end{figure*}
 sphere is calculated with Mie theory. Next, the band structure is calculated by solving the eigenvalue equation of the 3D structure with the full-wave solver mpGMM \cite{Beutel:21}. Results are shown in Figures~\ref{fig:GoldSpheresR1nmA205nm}\textbf{(a)} and \textbf{(b)} concerning the real and imaginary part of the propagation constant for a propagation direction along one of the principal axes. One observes that the band structure never approaches the edge of the Brillouin zone,
 \begin{figure*}
\centering
     \subfloat{
	\includegraphics[width=0.45\textwidth]{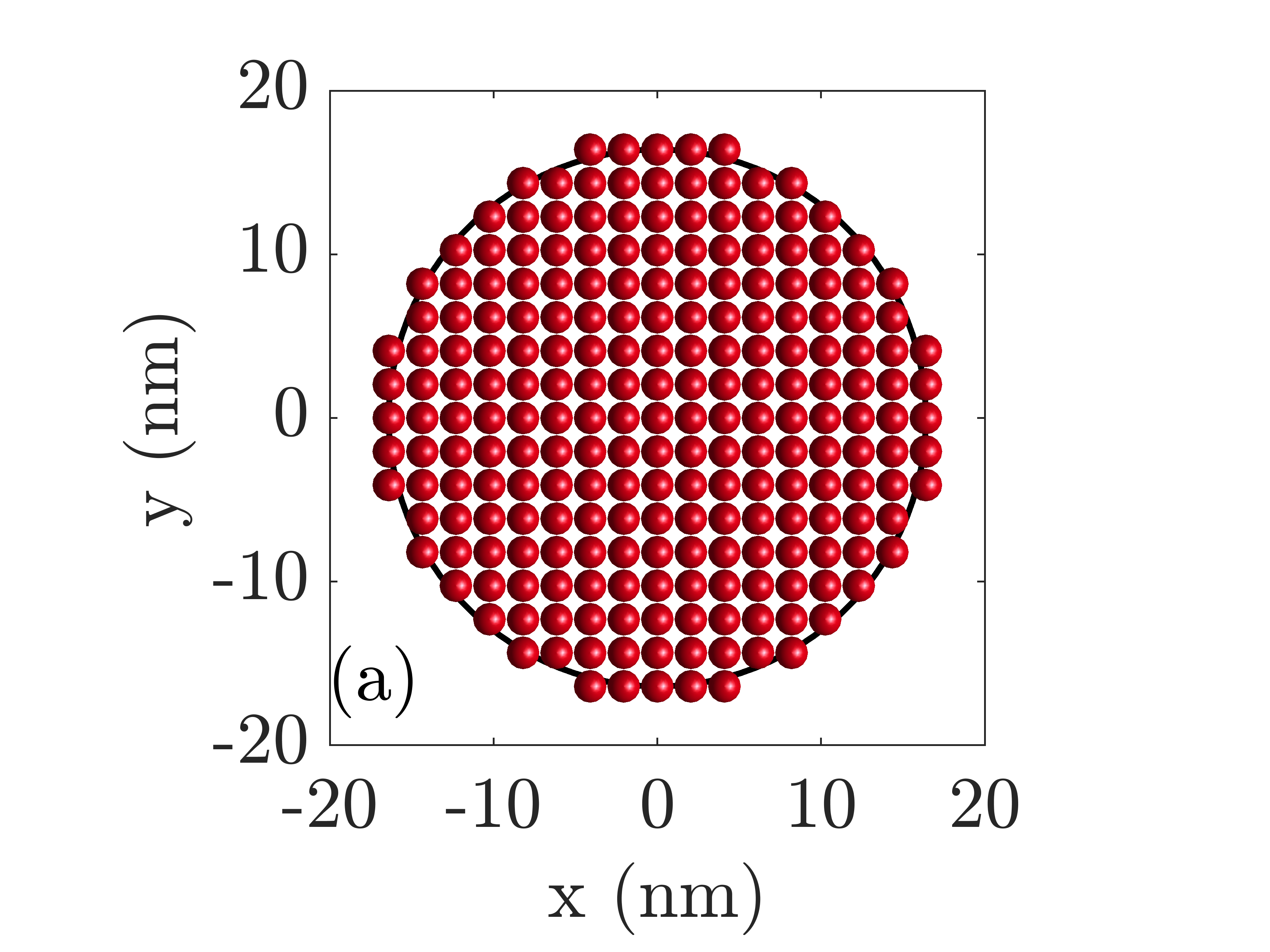}
	}
	\subfloat{
	\includegraphics[width=0.45\textwidth]{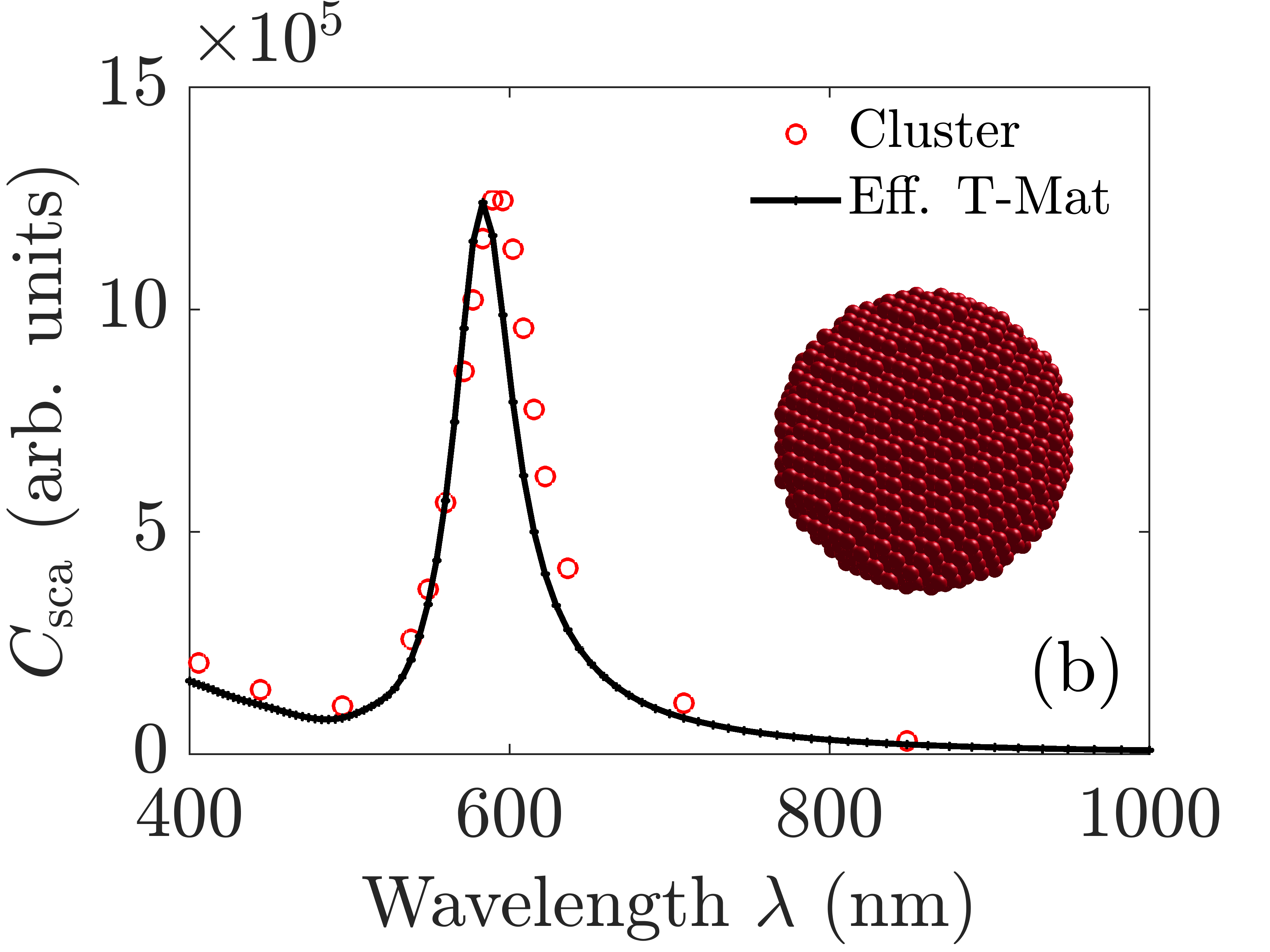}
	}

	\caption{An object of approximately spherical shape is made from the cubic lattice of gold spheres analyzed in Figure~\ref{fig:GoldSpheresR1nmA205nm}. Figure~\ref{fig:ScatCrossSec}\textbf{(a)} is an equatorial cross-cut of the arrangement. The black line is a circle with a radius of $R_{\mathrm{eff}}=8a=\SI{16.4}{\nano\meter}$. In \textbf{(b)}, the cross-section of the scattered plane waves is shown for both the cluster and an effective sphere with $R_{\mathrm{eff}}$ whose response is obtained with the effective material parameters derived from $\Teffdip$. The incident plane wave is linearly polarized. The exact scattering cross-section for the cluster is computed with CELES \cite{EGEL2017103}. Despite some differences because the cluster does not have a perfect spherical shape, the cross section calculated for the cluster agrees well with the one calculated with the effective material parameters. This confirms the expectation that the effective material parameters can be used independently of the shape of the target object.}
    \label{fig:ScatCrossSec}
	\end{figure*}
 where Bragg reflection would occur: The material is homogenizable, and $\Teff$ is then computed.
 
 Figure~\ref{fig:GoldSpheresR1nmA205nm}\textbf{(c)} shows the ratio between the amplitude of the electric dipolar entry of $\Teff$ and the corresponding value for the isolated sphere in $\Tcell$. Even though the response of an individual sphere is described by an electric dipolar polarizability to very good approximation ($N=1$), the lattice interactions involving up to the $N=5$ multipolar order modify the electric dipolar polarizability of the individual sphere significantly. In some frequencies, the amplitude of the electric dipolar entry of $\Teff$ is more than twice the corresponding value for the isolated sphere in $\Tcell$. The SI contains an extended discussion regarding the impact of the choice of $N$. Figure~S1 in the SI shows that such modifications are very much smaller when $N=1$: It is hence very important to incorporate higher multipolar orders in the calculation of $\Teff$. Since such lattice-induced effects affect the dipolar part of $\Teff$, they also impact the effective permittivity shown in Figure~\ref{fig:GoldSpheresR1nmA205nm}\textbf{(d)}, where we observe a very pronounced resonance close to \SI{800}{\nano\meter}. A much less pronounced resonance is also visible close to \SI{600}{\nano\meter}. We do not show the permeability as there is no notable magnetic resonance in this frequency range. The permeability is, however, included for calculating the responses of the slab and the sphere shown below. Figure S1 in the SI shows that $\tau\left(\Teff\right)<8\times10^{-5}$ in the whole frequency range, meaning that $\Teff$ is very much dominated by its dipolar part. We expect accurate results from the homogenized models of target objects of general shape.

We now study a particular geometry: a slab with a thickness of \SI{2.15}{\milli\meter}, corresponding to $2^{20}$ layers of gold spheres. The well-known Fresnel equations and standard boundary conditions are used to obtain the reflection coefficient for the slab using the effective material parameters. The obtained reflection coefficients are compared to the exact solutions for a slab made from the actual lattice of gold spheres, which are calculated with the full-waver solver mpGMM \cite{Beutel:21}. Due to the high absorption and large thickness, the transmission of the slab is zero. The illumination is a transverse magnetic (TM)-polarized plane wave with a wave vector in the XZ-plane, $\bm{\hat{k}}_{\mathrm{inc}}=[\sin(\theta),0,\cos(\theta)]^{\mathrm{T}}$. Two cases are considered: normal incidence ($\theta=0^{\circ}$) in Figure~\ref{fig:GoldSpheresR1nmA205nm}\textbf{(e)}, and oblique incidence with $\theta=75^{\circ}$ in Figure~\ref{fig:GoldSpheresR1nmA205nm}\textbf{(f)}. For normal incidence, TM polarization means a polarization along the negative x-direction. For oblique incidence, the polarization is correspondingly rotated [see e.g. Equation~(2.62) in Ref.~ \onlinecite{FerCorTHESIS}].

As the material is effectively isotropic, the xx-components of the effective permittivity and permeability are used to calculate the response of the effective slab. However, indeed, all the other components on the diagonal of the material tensors would be identical.  We observe that, as expected, the results obtained with mpGMM agree very well with the results calculated with the effective parameters.

We now consider a target object with a different shape. The same discrete gold-spheres-in-cubic-lattice material is used to form a cluster with an approximately spherical shape of radius $R_{\mathrm{eff}}=8a=\SI{16.4}{\nano\meter}$ [Figure~\ref{fig:ScatCrossSec}\textbf{(a)}, inset in  Figure~\ref{fig:ScatCrossSec}\textbf{(b)}]. In Figure~\ref{fig:ScatCrossSec}\textbf{(b)}, we compare the scattering cross sections calculated in two different ways: using CELES \cite{EGEL2017103} for computing the exact solution for the cluster (see Methods), and using Mie theory for a homogeneous effective sphere of radius $R_{\mathrm{eff}}$ with the effective material parameters obtained with $\Teffdip$. We observe that the scattering cross sections agree well, confirming the expectation that the effective material parameters can be used independently of the shape of the target object. The resonance we see corresponds to a localized plasmon-polariton excited in the sphere at the wavelength where the effective permittivity in very good approximation satisfies the Fr\"ohlich condition. The differences observed in Figure~\ref{fig:ScatCrossSec}\textbf{(b)} can be attributed to the fact that the cluster does not have a perfect spherical shape. We also note that, in order to apply any homogenization technique, a cluster should have a sufficiently large number of unit cells, and some inaccuracies could also originate from having a finite number of spheres in the cluster.

\section{Cut-plate pairs in a cubic lattice}
\begin{figure*}[t!]
\centering
\subfloat{
\includegraphics[width=0.45\textwidth]{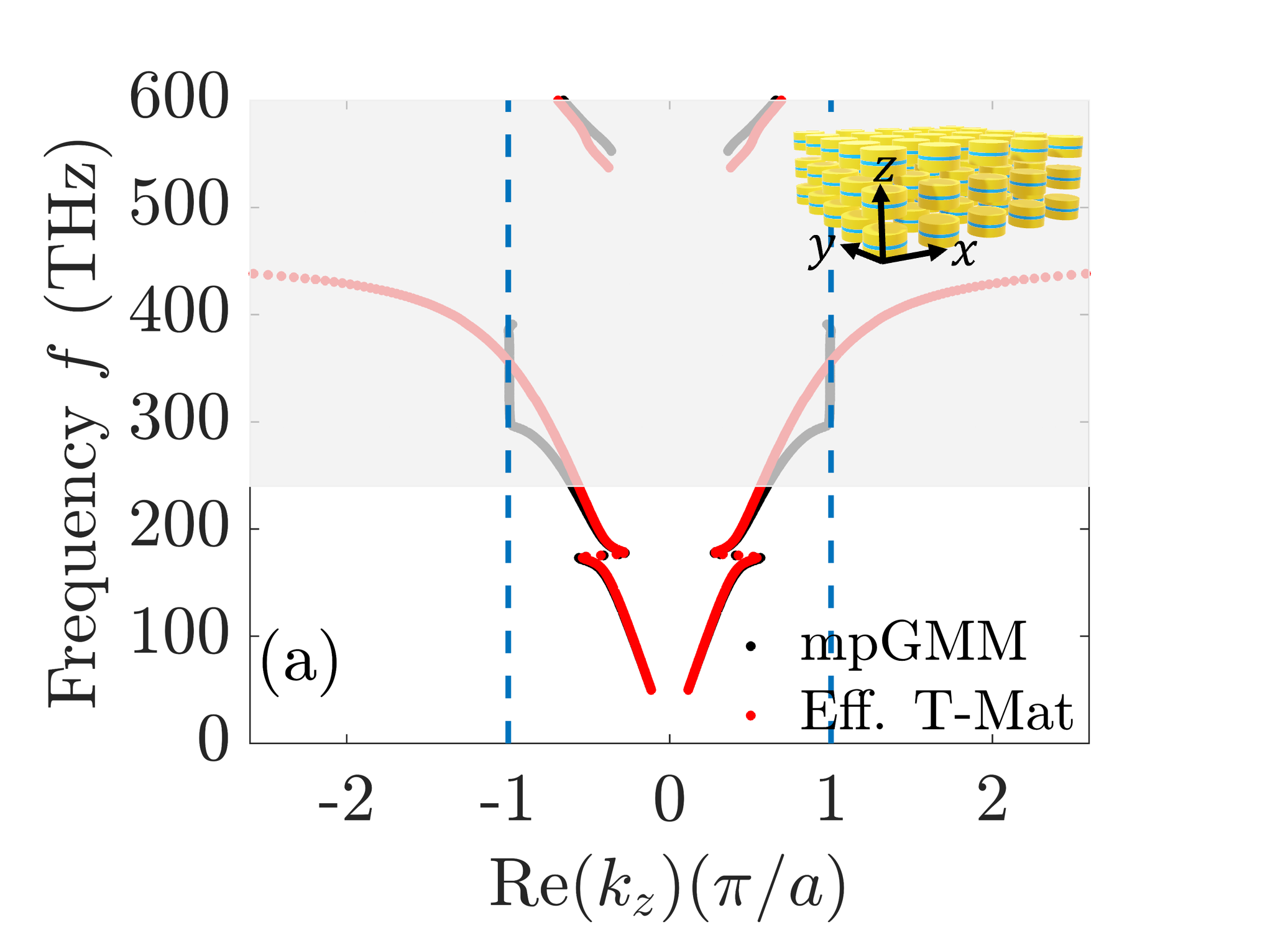}
}
\subfloat{
\includegraphics[width=0.45\textwidth]{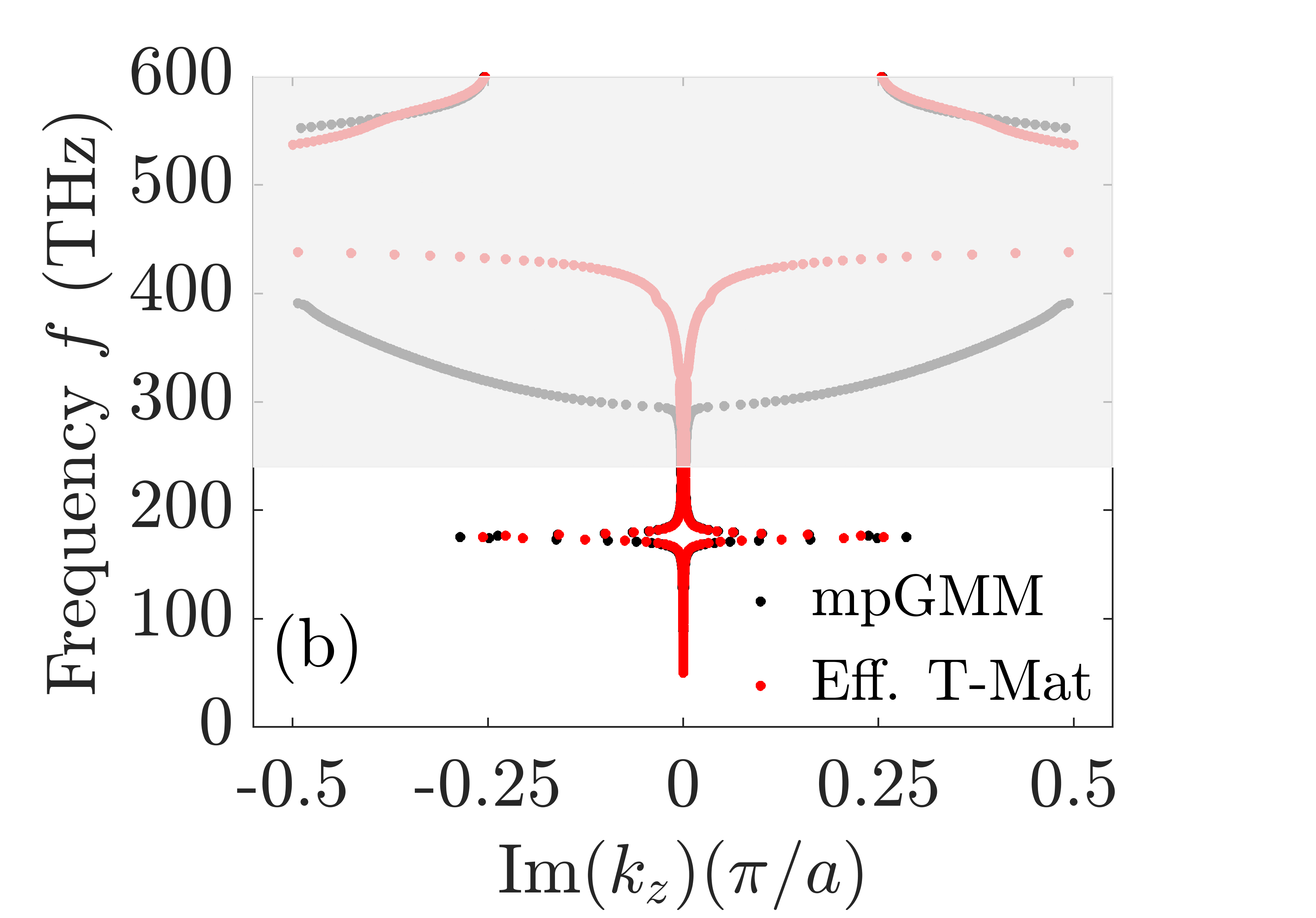}
}\\
   \subfloat{
	\includegraphics[width=0.45\textwidth]{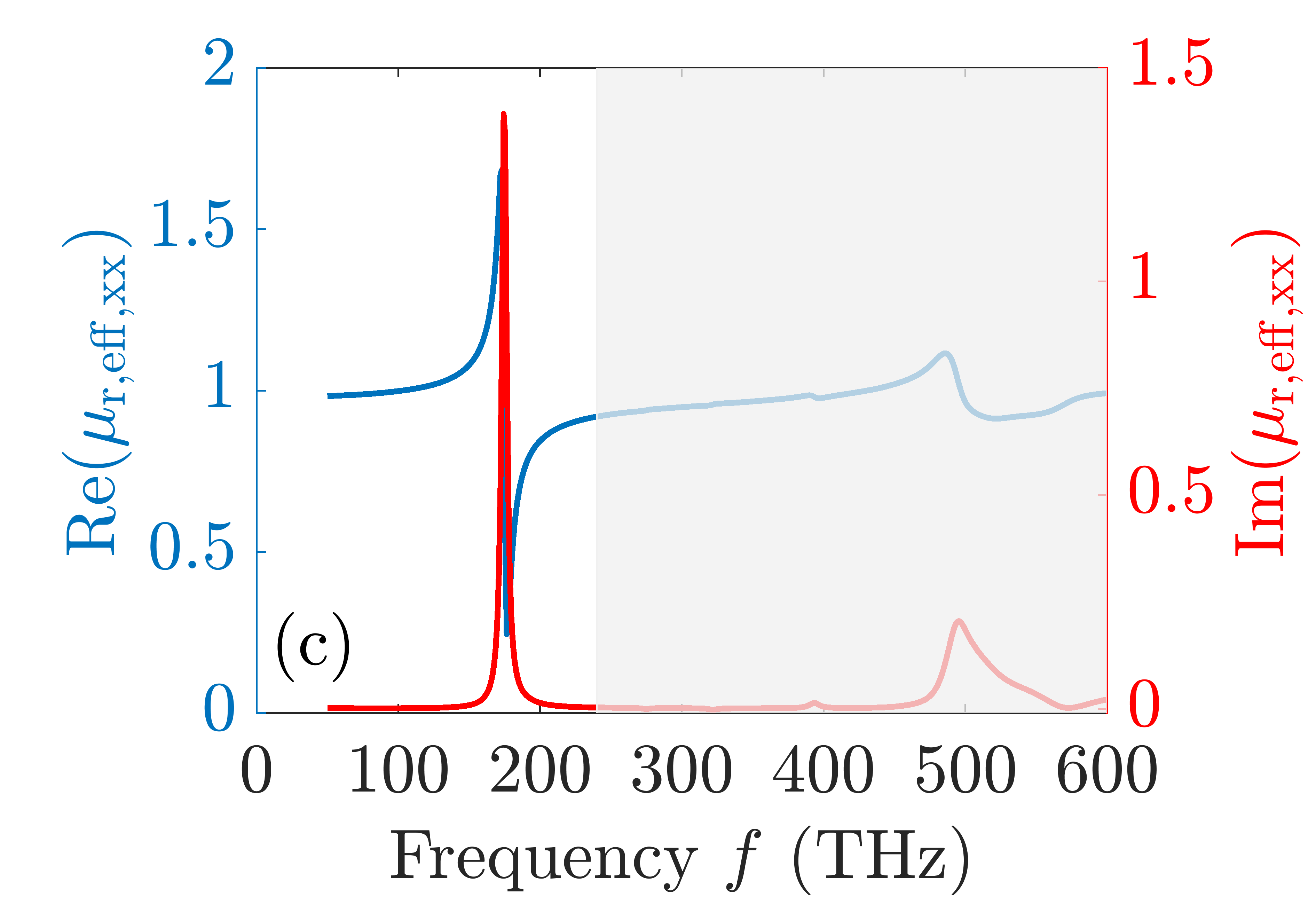}
	}
    \subfloat{
	\includegraphics[width=0.45\textwidth]{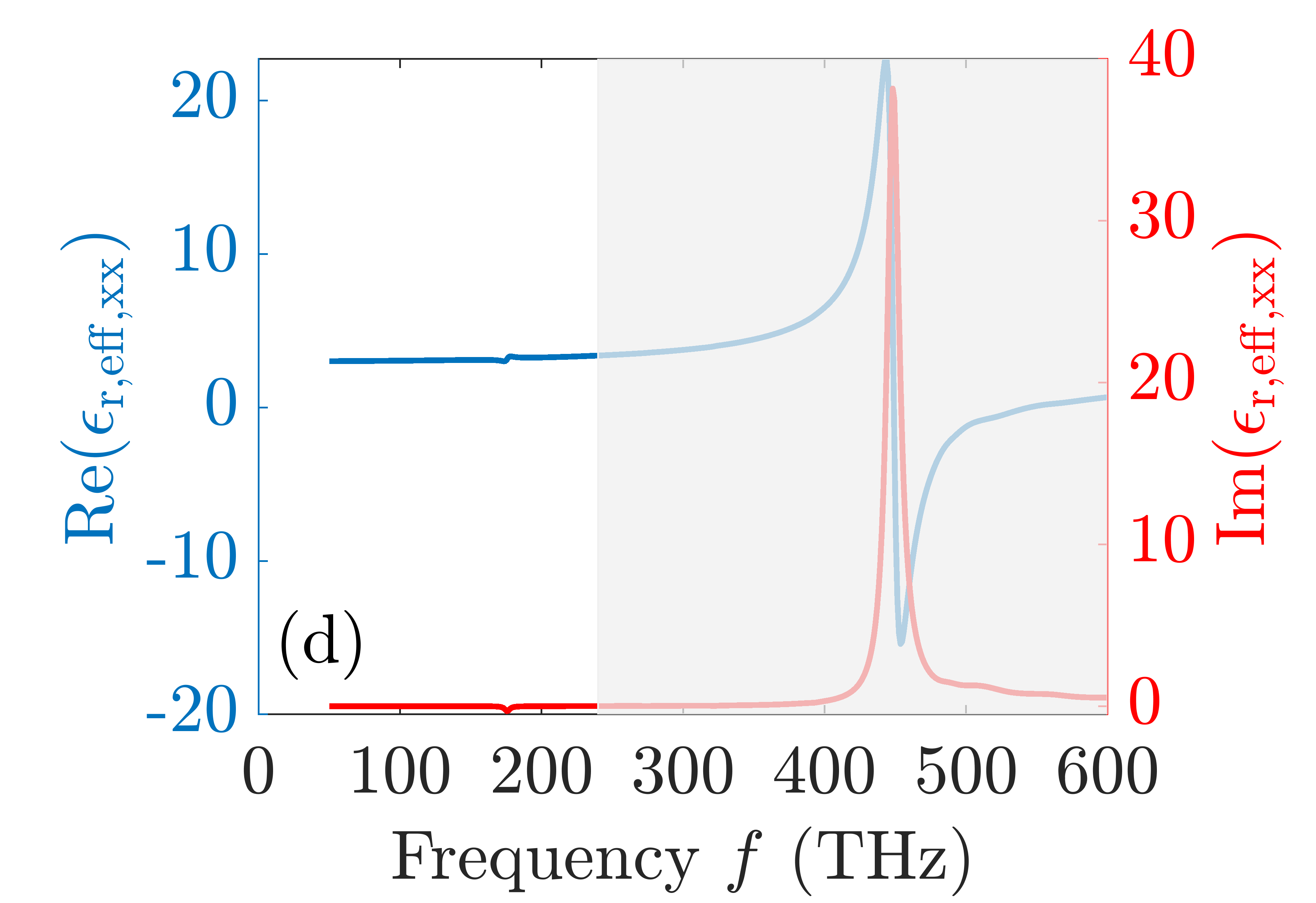}
	}\\
	\subfloat{
	\includegraphics[width=0.45\textwidth]{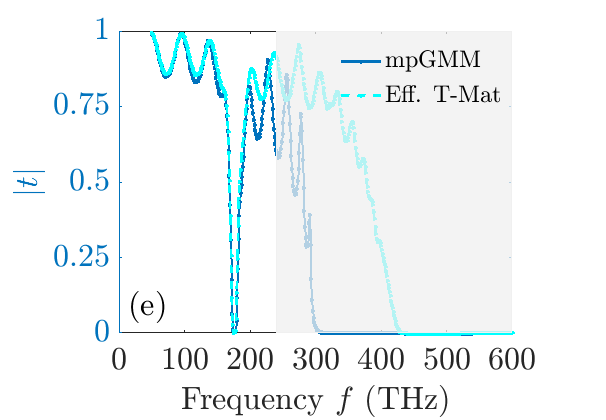}
	}
    \subfloat{
	\includegraphics[width=0.45\textwidth]{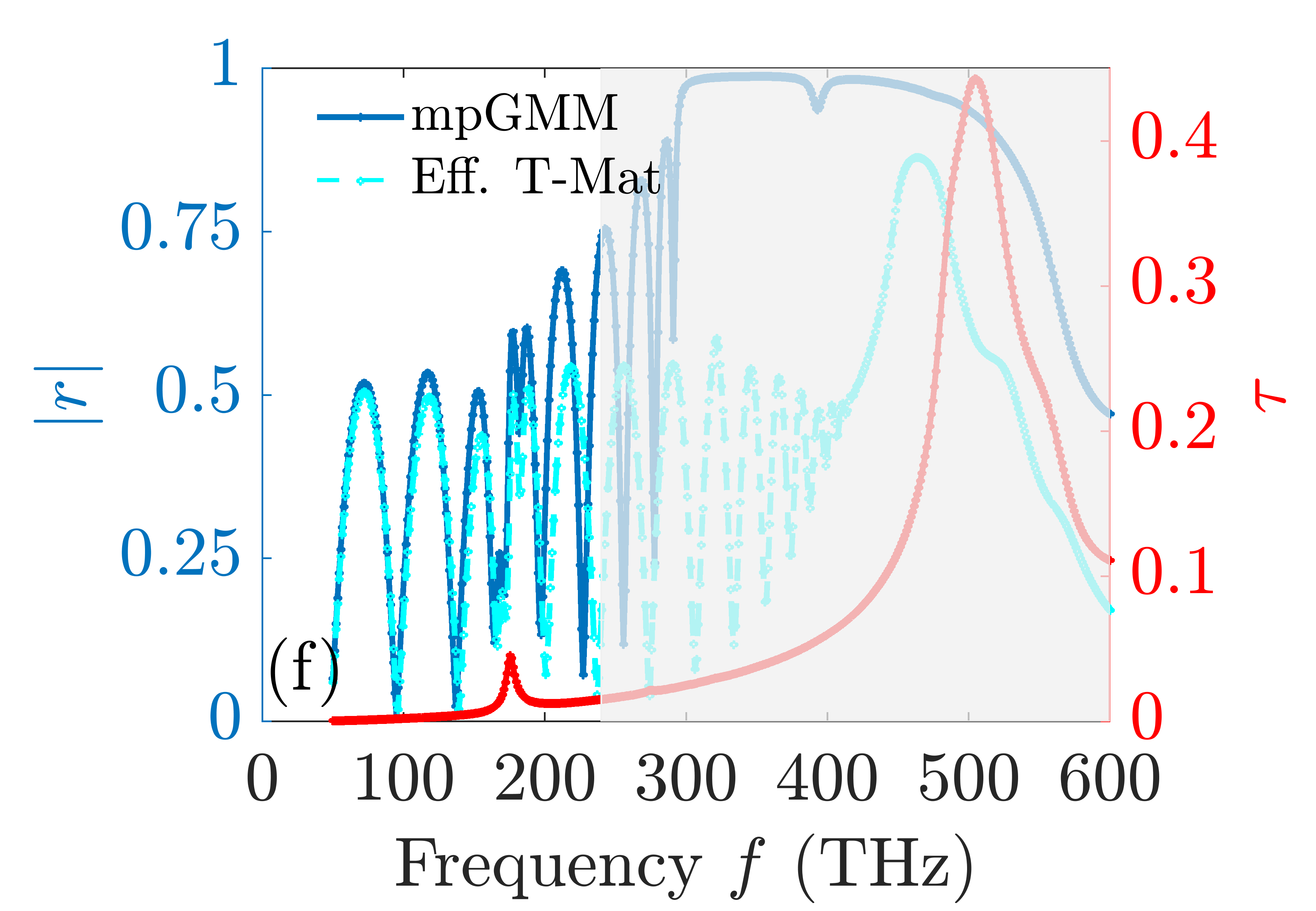}
	}
	
	\caption{\textbf{(a),(b)} Band structure of a 3D lattice made from cut-plate pairs calculated with mpGMM for real and imaginary $k_z$ together with the dispersion relation $k_z=\frac{2\pi f}{c_0}\sqrt{\varepsilon_{\mathrm{r,eff,xx}}\mu_{\mathrm{r,eff,xx}}}$ obtained from the effective homogenized material parameters. A band gap appears between 300\,THz and 550\,THz due to Bragg reflection at the edge of the Brillouin zone. This bends the dispersion bands from approximately 240\,THz on, which is marked by the gray shade indicating that the light-matter interaction in the actual material can then not be reliably modeled by a homogeneous medium. Permeability \textbf{(c)} and permittivity \textbf{(d)} of the material computed with the effective T-matrix. Absolute value of the transmission \textbf{(e)}, and reflection \textbf{(f)} coefficients of a slab under normal incidence. The dark blue lines correspond to the exact solution of nine layers of the cut-plate pairs stacked in the $z$ direction. The light blue lines correspond to the solution of a homogeneous slab assuming a thickness of \SI{1800}{\nano\meter}.} 
    \label{fig:HomCPP}	
	\end{figure*}

In order for us to explore the limits of the homogenization method, we now consider a more extreme photonic material made from cylindrically-shaped cut-plate pairs with a radius of $R=\SI{90}{\nano\meter}$ and a cubic lattice constant of $a=\SI{200}{\nano\meter}$, embedded in vacuum. The cut-plate pairs consist of two gold layers with a thickness of $d_{\mathrm{Au}}=\SI{30}{\nano\meter}$ separated by an insulator with a thickness of $d_{\mathrm{Ins}}=\SI{5}{\nano\meter}$ and made from a material with an isotropic, nondispersive dielectric characterized by ${\varepsilon}_{\mathrm{r,Ins}}=2.25$ \cite{PhysRevB.89.155125}. The material parameters of gold are taken from \cite{Babar:15}. The T-matrix of a cut-plate pair is calculated with JCMsuite \cite{PhysRevB.99.045406,JCM}. In Figures~\ref{fig:HomCPP}\textbf{(a)} and \textbf{(b)}, the band structure of the material is shown. We observe a band gap between 300\,THz and 550\,THz in which no propagation occurs. This Bragg gap is caused by reflection at the edge of the Brillouin zone. 

 At approximately 175\,THz, far away from the edge of the Brillouin zone, a resonance occurs, bending the dispersion relation. This resonance can be related to a multipolar resonance. In the range of this resonance, the influence of the Bragg gap and higher resonances can be neglected. Starting approximately at 200\,THz, the influence of the multipole resonance vanishes. For frequencies higher than the turning point of the band at approximately 240\,THz, the influence of the Bragg gap bends the dispersion bands, and homogenization becomes unreliable. This is marked by the gray shading in the figures.
 
In \cite{PhysRevB.89.155125}, it is shown that the single cut-plate pair has a magnetic dipole resonance at low frequencies and an electric dipole resonance at high frequencies. In Figures~\ref{fig:HomCPP}\textbf{(c)} and \textbf{(d)}, we observe, indeed, a distinct resonance on the magnetic permeability for a frequency below the \SI{240}{\tera\hertz} limit. Above such limit, we observe a resonance of the electric permittivity and an additional resonance of the magnetic permeability due to the electric quadrupole coupling to the magnetic dipole \cite{rahimzadegan2021comprehensive}. The first $N=7$ multipolar orders were included in the calculations of $\Teff$. The $\tau\left(\Teff\right)$ metric of \Eq{eq:diptest} is shown in Figure~\ref{fig:HomCPP} \textbf{(f)}, where we see that $\tau\left(\Teff\right)$ is more than two orders of magnitude higher than in the previous example, including a peak value of 0.045 in the homogenizable frequency range, and an increasing trend towards a very large peak well beyond the \SI{240}{\tera\hertz} limit.

In the following, we consider a slab of nine layers of the cut-plate pairs in the $z$ direction. For the homogeneous model, we take $\SI{1800}{\nano\meter}$ as the thickness of the slab in the $z$ direction. The transmission and reflection coefficients of the non-homogeneous slab for normal incidence are calculated with mpGMM, and those for the homogeneous slab are calculated with the effective parameters from Equation~(\ref{eq:EffParam}). As the principal axes of the material coincide with the Cartesian axes, the light propagates along the z-axis, and the material is the same in x- and y-direction, we use the xx-component of both the permittivity and permeability to calculate the response of the effective homogeneous slab. Figures~\ref{fig:HomCPP}\textbf{(e)} and \textbf{(f)} contain the results.  

We observe that when both criteria are satisfied, the results from the homogeneous slab match very well those of the discrete structure. Both results show the pronounced effect of the magnetic dipole resonance around \SI{175}{\tera\hertz}. The results start to disagree after \SI{200}{\tera\hertz} due to the influence of the Bragg resonance. We also see in Figure~\ref{fig:HomCPP}\textbf{(a)}, that from 200\,THz on the dispersion relation calculated with the effective parameters differs from the band structure calculated with mpGMM. 

\section{A bi-anisotropic and chiral molecular material}
\begin{figure*}
\centering

\subfloat{
	\includegraphics[width=0.35\textwidth]{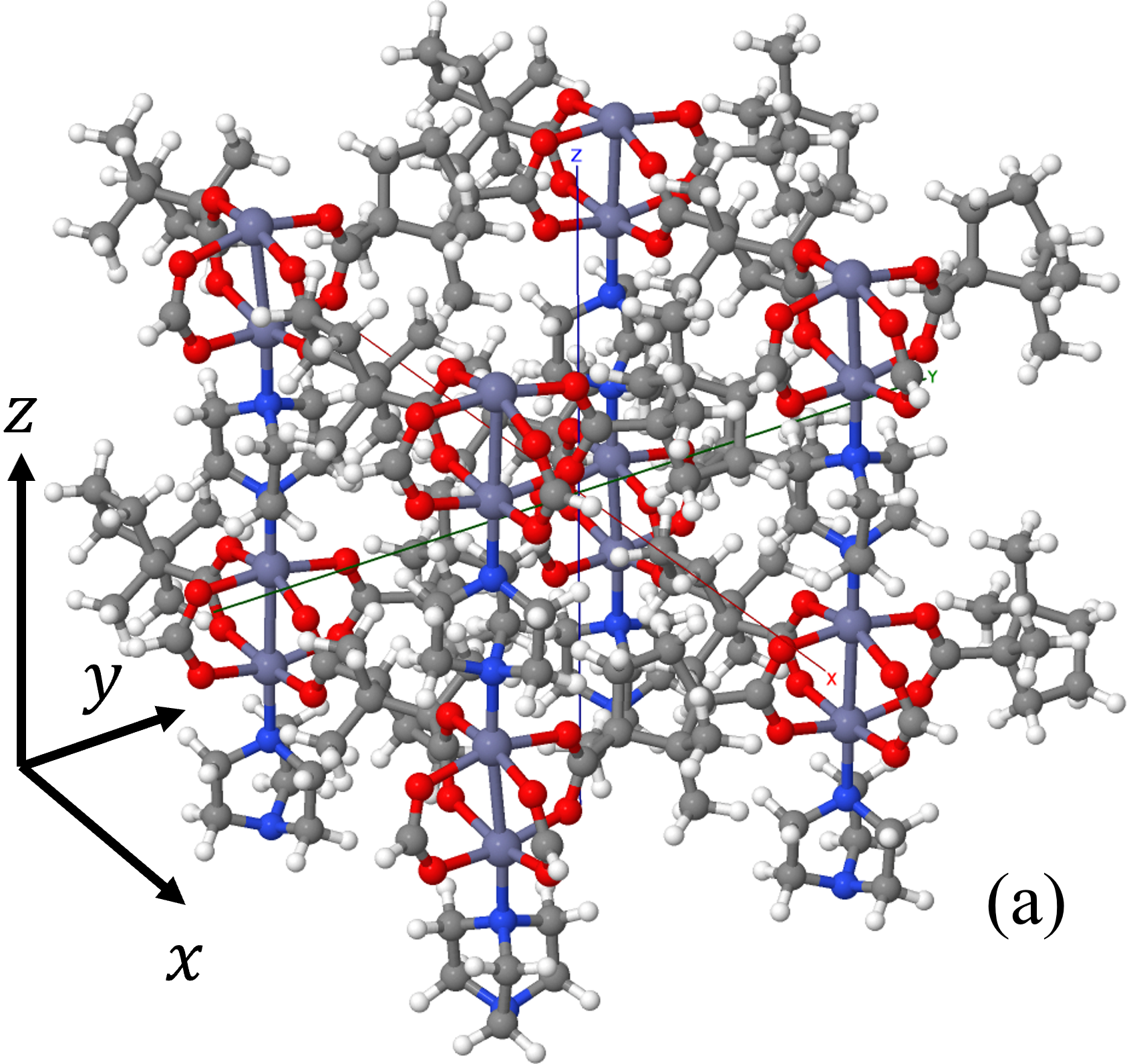}
	}
	\hspace{1cm}
	\subfloat{
	\includegraphics[width=0.45\textwidth]{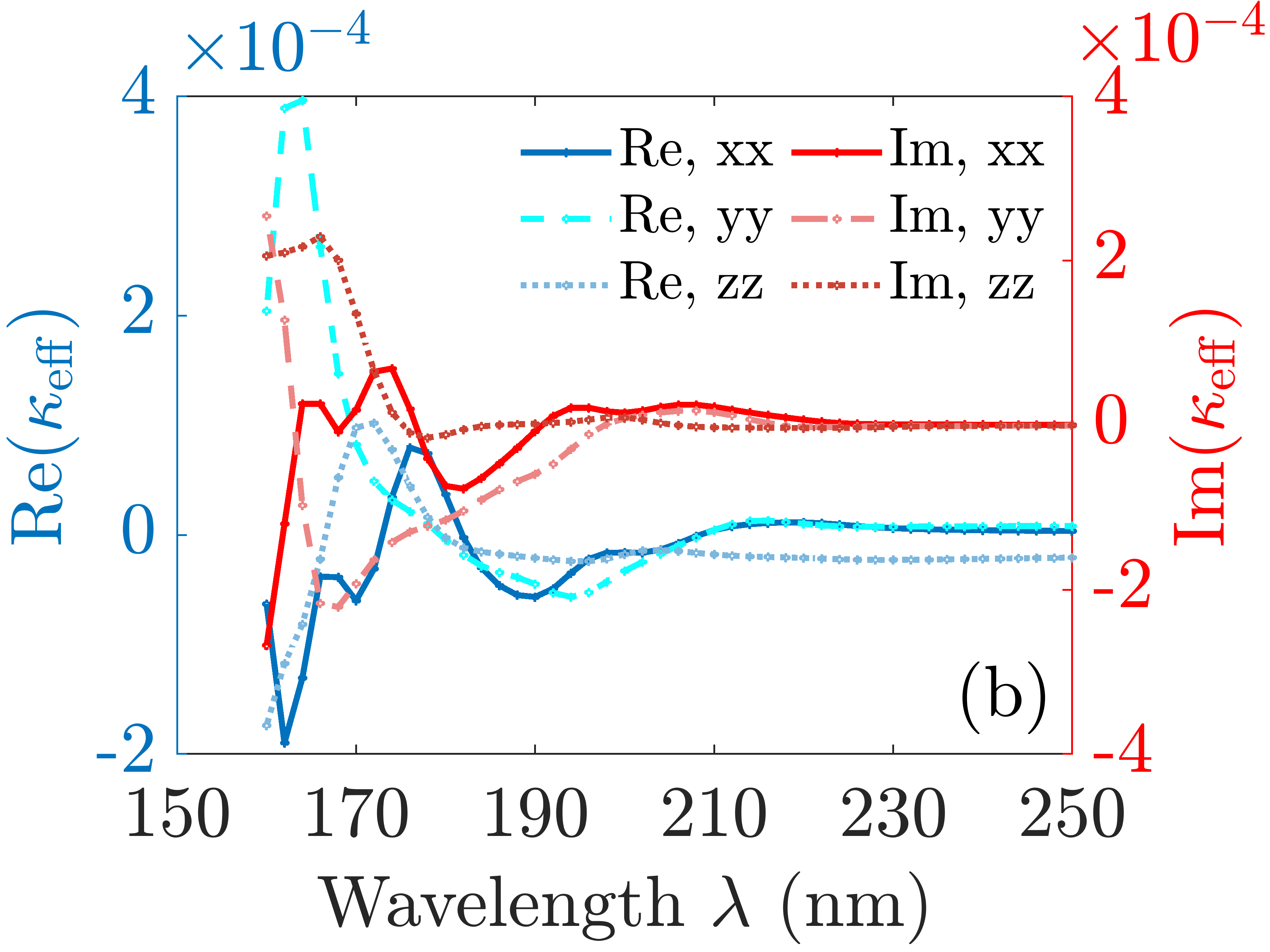}
	}\\
   \subfloat{
	\includegraphics[width=0.45\textwidth]{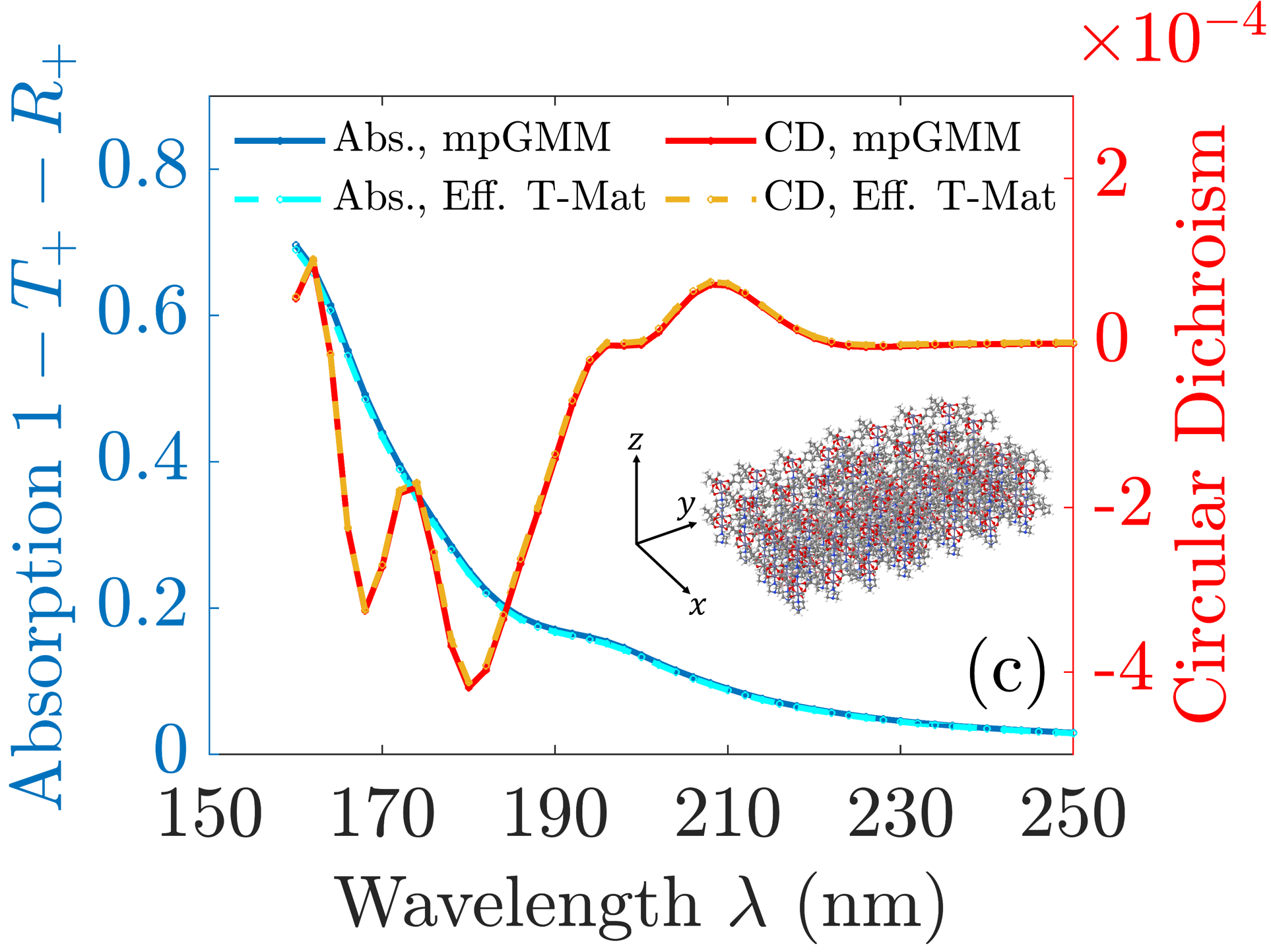}
	}
	\subfloat{
	\includegraphics[width=0.45\textwidth]{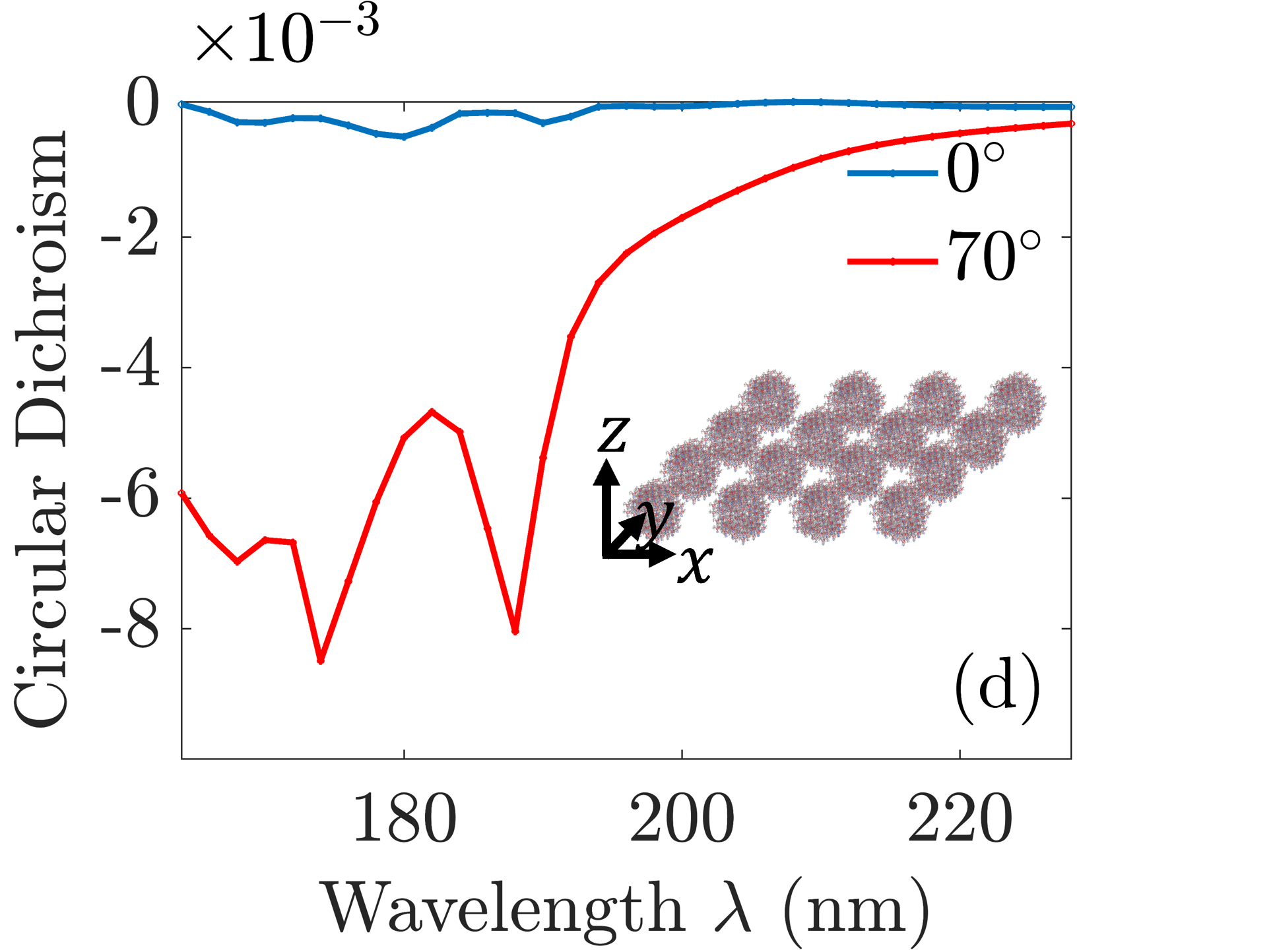}
	}
	
	\caption{In a Zn-L-camphoric acid-dabco SURMOF, the T-matrix of the individual isolated cell \textbf{(a)} is computed using TD-DFT. \textbf{(b)} Real and imaginary parts of the effective chirality. We observe that the chirality is anisotropic. \textbf{(c)} Comparison of the absorption of an incident left-circularly polarized plane wave and the circular dichroism of a $\SI{77}{\nano\meter}$ thick slab computed with the effective material parameters, and directly with mpGMM which considers the discrete lattice explicitly \cite{SURMOFCavity}. The two results match perfectly. \textbf{(d)} Circular dichroism of a planar array of spheres made from the SURMOF material for normal and oblique incidence. The spheres have a radius $R=\SI{72}{\nano\meter}$ and the lattice constant of the square lattice is $a=\SI{162}{\nano\meter}$.}
    \label{fig:ChiralSURMOF}	
	\end{figure*}

We now demonstrate the wide range of applicability of our homogenization method by considering a material from a completely different class: a Zn-L-camphoric acid-dabco SURMOF, which has a chiral bi-anisotropic structure. The SURMOF consists of the chiral L-camphoric acid linker molecules, which build a layer together with Zn paddle wheels [Figure~\ref{fig:ChiralSURMOF}\textbf{(a)}]. The layers forming the SURMOF are connected by dabco pillar linkers. The band diagrams in Figure~S2 show that the molecular material is homogenizable in the considered frequency range. The T-matrix of the unit cell is computed using TD-DFT \cite{SURMOFCavity}. The lattice constants are in x- and y-direction $a_1=a_2=\SI{2.079}{\nano\meter}$, and in z-direction $a_3=\SI{1.922}{\nano\meter}$. At optical frequencies, only the dipolar response needs to be considered because the unit cells have sizes with linear dimensions of the order of \SI{2}{\nano\meter}. The criterium $\tau\left(\Teff\right)\rightarrow 0$ is always satisfied. 

In the homogeneous model of \Eq{eq:ConsRel}, $\bm{\kappa}_{\mathrm{eff}}$ is responsible for the chiro-optical effects such as circular dichroism (CD), i.e., the differential absorption of left- and right-hand polarized light. Figure~\ref{fig:ChiralSURMOF}\textbf{(b)} shows the real and imaginary parts of $\bm{\kappa}_{\mathrm{eff}}$. We observe that they are different along the directions of the different lattice vectors of the structure, which implies that $\kappa_{\mathrm{eff}}$ is anisotropic in this material. We focus on the circular dichroism and, therefore, on the chirality. The permittivity and permeability are not shown as they are of minor importance in this context. As the material is reciprocal, $\bm{\gamma}_{\mathrm{eff}}=-\bm{\kappa}^{\mathrm{T}}_{\mathrm{eff}}$ \cite{Kristensson2016}, and, therefore, one can additionally describe magneto-electric coupling solely with $\bm{\kappa}_{\mathrm{eff}}$. 

We first consider a slab with a thickness of \SI{77}{\nano\meter}. Inserting the effective material parameters into COMSOL, we compute, for normal incidence, the absorption spectrum for a left-handed circularly polarized plane wave and the circular dichroism of the film. We define the circular dichroism as $\mathrm{CD}=\frac{A_+-A_-}{2}$, where $A_+$ and $A_-$ are the absorption of left-handed and right-handed circularly polarized plane waves, respectively. We observe in Figure~\ref{fig:ChiralSURMOF}\textbf{(c)} that the results obtained with the homogeneous model match perfectly those obtained with mpGMM, which explicitly considers the discrete SURMOF lattice \cite{SURMOFCavity}. 

The second example is a planar array of spheres embedded in vacuum. The spheres are made from the SURMOF material, have a radius of $R=\SI{72}{\nano\meter}$, and are arranged in a square lattice with constant $a=\SI{162}{\nano\meter}$. In Figure~\ref{fig:ChiralSURMOF}\textbf{(d)}, the CD calculated with COMSOL is displayed for normal incidence and oblique incidence at $70$ degrees. We observe that the circular dichroism is much more pronounced for oblique incidence. This prediction is only possible and trustworthy due to the confidence that can be placed in the effective material parameters when the two homogenization criteria are met. This example illustrates the use of the constitutive relations in generic Maxwell solvers, allowing the simulation of target objects of general shape. Moreover, planar systems containing different lattices \cite{Oldenburg2016} can now be efficiently simulated by codes that combine the T-matrix and Ewald summation methods. The simulation is possible once all the lattices, or all but one of them are homogenized.

We highlight that the T-matrices of the unit cells of molecular materials, and hence ultimately the material parameters, are obtained from {\em ab initio} quantum-chemical computational methods \cite{Fernandez-Corbaton:2020}, such as TD-DFT.

\section{Conclusion and outlook}
We have introduced a method for homogenizing artificial materials made by three-dimensional lattices of electromagnetic scatterers. The starting point for homogenization is the non-spatially dispersive yet exact response of the discrete material, including all lattice interactions. The material parameters of the homogeneous effective medium are determined from the dipolar part of such exact response without involving any particular shape of a target object. This truncation to dipolar order is the only physically significant approximation in the method. The resulting bi-anisotropic constitutive relations and boundary conditions are the ones implemented in standard Maxwell solvers. We have shown that, independently of the shape of the target objects, the electromagnetic response of finite objects made from the actual 3D lattice of scatterers is very well predicted by the corresponding effective homogeneous models, provided that two criteria are met. One is that light should not experience any explicit lattice effects such as Bragg resonances. This criterium can be assessed using the band structure of the discrete material and determines whether the material is homogenizable at all, independently of the homogenization method. The other criterium is that the difference between the exact description of the discrete material and its dipolar part should be small. Both tests are independent of the shape of any target object. 

We are confident that the method will be helpful for the computer-aided design of photonic devices containing artificial materials and for interpreting experimental measurements. In particular, the method is suitable for objects fabricated by three-dimensional laser printing, and/or containing structured molecular materials.

A plausible extension of the method would include the quadrupolar orders of the exact response in the homogeneous model, thereby extending the range of applicability to materials that are homogenizable in principle, but where the contribution of orders higher than the dipole cannot be neglected.

\section{Methods}
\subsection{Analytical derivation of $\Teff$}

Let us fix the wavenumber $k=\sqrt{\bm{k}\cdot\bm{k}}$  and remove it from the notation. Let $\mathbf{\hat{\tilde{T}}}_{\mathrm{eff}}(\bm{\hat{k}})$ be the operator corresponding to the $\bm{k}$-dependent effective T-matrix in Equation~(\ref{eq:Teff}) so that

	\begin{align}\label{eq:ps}
        \mathbf{\hat{\tilde{T}}}_{\mathrm{eff}}(\bm{\hat{k}})\vert \bm{\hat{k}},\lambda\rangle=\sum_{\tilde{j},\tilde{m},\tilde{\lambda}}\bm{\tilde{p}}^{\tilde{j},\tilde{m},\tilde{\lambda}}(\bm{\hat{k}},\lambda)\vert \tilde{j},\tilde{m},\tilde{\lambda}\rangle ,
    \end{align}
    
   \noindent where $\vert\bm{\hat{k}},\lambda\rangle$ represents an incident plane wave with well-defined polarization handedness (helicity) $\lambda=\pm1$ and propagation direction $\bm{\hat{k}}$, and $\bm{\tilde{p}}^{\tilde{j},\tilde{m},\tilde{\lambda}}(\bm{\hat{k}},\lambda)$ are the coefficients of the far-field scattered wave expanded in vector spherical harmonics $\vert \tilde{j},\tilde{m},\tilde{\lambda}\rangle$. In $\vert j,m,\lambda\rangle$, $j=1$ corresponds to the dipolar order, $j=2$ to the quadrupolar order, etc ..., and $m=-j,\ldots,j$ is the angular momentum of the spherical wave along the $z$ axis. Furthermore, let 
   
    \begin{align}\label{eq:mpols}
    \begin{split}
		\vert j,m,\lambda\rangle &=\int \mathrm{d}\bm{\hat{k}}  \beta^{j,m,\lambda}(\bm{\hat{k}})\vert \bm{\hat{k}},\lambda\rangle\\&=\int_{0}^\pi \mathrm{d}{\theta}_{\bm{\hat{k}}}\sin{\theta}_{\bm{\hat{k}}}\int_{-\pi}^\pi\mathrm{d}\varphi_{\bm{\hat{k}}}\beta^{j,m,\lambda}(\bm{\hat{k}})\vert \bm{\hat{k}},\lambda\rangle
   \end{split}
   \end{align}
	be the expansion of a vector spherical wave $\vert j,m,\lambda\rangle$ in plane waves $\vert \bm{\hat{k}},\lambda\rangle$, where ${\theta}_{\bm{\hat{k}}}=\arccos\left(k_z/k\right)$, and $\varphi_{\bm{\hat{k}}}=\arctan(k_y,k_x)$.
	
	The expansion coefficients $\beta^{j,m,\lambda}(\bm{\hat{k}})$ are defined as
	
    \begin{align}
		\beta^{j,m,\lambda}(\bm{\hat{k}})=\frac{\gamma_{j,m}}{4\pi\mathrm{i}^{j+1}}\left(\frac{m}{\sin{\theta_{\bm{\hat{k}}}}}P^j_m(\cos{\theta_{\bm{\hat{k}}}})+\lambda\frac{\partial}{\partial\theta_{\bm{\hat{k}}}} P^j_m(\cos{\theta_{\bm{\hat{k}}}})\right)
    \end{align}
    
\noindent with, see Equations (S3c,S3d) from \cite{Beutel:21}, 

\begin{align}
	\gamma_{j,m}&=\mathrm{i}\sqrt{\frac{2j+1}{4\pi}\frac{(j-m)!}{j(j+1)(j+m)!}},
\end{align}
and $P^j_m(\cos{\theta_{\bm{\hat{k}}}})$ are the associated Legendre polynomials .

Then, the matrix elements of the {\em direction-independent} effective T-matrix in the multipolar basis $\Teff$, can be derived as

    \begin{align}\label{eq:IntTeff}
        &\langle \bar{\lambda},\bar{m},\bar{j}\vert \mathbf{\hat{T}}_{\mathrm{eff}} \vert j,m,\lambda\rangle
        =\int \mathrm{d}\bm{\hat{k}}  \beta^{j,m,\lambda}(\bm{\hat{k}}) \langle \bar{\lambda},\bar{m},\bar{j}\vert\mathbf{\hat{\tilde{T}}}_{\mathrm{eff}}(\bm{\hat{k}})\vert\bm{\hat{k}},\lambda\rangle\nonumber\\ 
		&=\sum_{\tilde{j},\tilde{m},\tilde{\lambda}}\int \mathrm{d}\bm{\hat{k}}  \beta^{j,m,\lambda}(\bm{\hat{k}}) \bm{\tilde{p}}^{\tilde{j},\tilde{m},\tilde{\lambda}}(\bm{\hat{k}},\lambda) \langle \bar{\lambda},\bar{m},\bar{j}\vert \tilde{j},\tilde{m},\tilde{\lambda}\rangle\\
        &= \int \mathrm{d}\bm{\hat{k}}  \beta^{j,m,\lambda}(\bm{\hat{k}})\bm{\tilde{p}}^{\bar{j},\bar{m},\bar{\lambda}}(\bm{\hat{k}},\lambda),\nonumber
    \end{align}
where the first equality follows from \Eq{eq:mpols} and the key imposition that $\Teff$ shall respond to an incident plane wave with a specific propagation direction $\bm{\hat{k}}$ as $\mathbf{\tilde{T}}_{\mathrm{eff}}(\bm{k})$ responds. The second follows from \Eq{eq:ps}, and the third from the orthonormality of the multipolar fields $\langle \bar{\lambda},\bar{m},\bar{j}\vert \tilde{j},\tilde{m},\tilde{\lambda}\rangle=\delta_{\bar{j},\tilde{j}}\delta_{\bar{m},\tilde{m}}\delta_{\bar{\lambda},\tilde{\lambda}}$. 

The result in the last line of \Eq{eq:IntTeff} features an integral over the sphere of $\bm{\hat{k}}$ directions. In practice, a finite number of directions must be selected, and $\bm{\tilde{p}}^{\bar{j},\bar{m},\bar{\lambda}}(\bm{\hat{k}},\lambda)$ computed for each direction. A particularly useful method for selecting equally-spaced points on a sphere, which transforms the integral into a Riemann sum of equally weighted terms can be found in \cite{partSamp}. We show in Figure~S3 of the SI for the examples of cut-plate pairs that the normalized difference between the effective T-matrices calculated with Equations~(\ref{eq:ModTeff}) and (\ref{eq:IntTeff}) is negligibly small.

\subsection{Deriving effective material parameters from $\Teff$} 
We consider scatterers in a periodic lattice, surrounded by an achiral non-magnetic host medium with permittivity $\varepsilon_{\mathrm{h}}=\varepsilon_{\mathrm{r,h}}\varepsilon_0$ and permeability $\mu_{\mathrm{h}}=\mu_0$. Here, $\varepsilon_0$ and $\mu_0$ are the vacuum permittivity and permeability, respectively. In the following, we omit the frequency $\omega$ of the incident wave as argument. All quantities besides purely geometric factors are, however, frequency-dependent, and $\exp\left(-\ii\omega t\right)$ factors are suppressed from the notation.

Externally applied electric $\bm{E}_{\mathrm{ext}}$ and magnetic $\bm{H}_{\mathrm{ext}}$ fields induce effective electric $\bm{P}_{\mathrm{eff,e}}$ and magnetic polarizations $\bm{P}_{\mathrm{eff,m}}$ in a scatterer in the lattice. We assume, without loss of generality, that this scatterer is placed at the origin of the 3D lattice. The effective polarizations of the scatterer are determined from the effective T-matrix ($\Teff$) of the lattice. Restricting our consideration to $\Teffdip$, the dipolar part of $\Teff$, the polarizations can be written as

\begin{align}\label{eq:Pol}
\begin{pmatrix}
	\bm{P}_{\mathrm{eff,e}}\\
	\bm{P}_{\mathrm{eff,m}}
\end{pmatrix}
=nq\begin{pmatrix}
\mathbf{T}^{j,j'=1,1}_{\mathrm{eff,EE,cart}}&\mathrm{i}Z_{\mathrm{h}}\mathbf{T}^{j,j'=1,1}_{\mathrm{eff,EM,cart}}\\
\mathrm{-i}Z_{\mathrm{h}}\mathbf{T}^{j,j'=1,1}_{\mathrm{eff,ME,cart}}&Z_{\mathrm{h}}^2\mathbf{T}^{j,j'=1,1}_{\mathrm{eff,MM,cart}}\end{pmatrix}
\begin{pmatrix}
\bm{E}_{\mathrm{ext}}\\
\bm{H}_{\mathrm{ext}}
\end{pmatrix},
\end{align}

\noindent where $n$ is the concentration of the scatterers per unit cell, $Z_{\mathrm{h}}=\sqrt{\mu_{0}/\varepsilon_{\mathrm{h}}}$ the wave impedance of the host medium, $q=\frac{-\mathrm{i}6\pi}{c_{\mathrm{h}}Z_{\mathrm{h}}k_{\mathrm{h}}^3}$ \cite{Fernandez-Corbaton:2020}, $c_{\mathrm{h}}=1/\sqrt{{\varepsilon}_{\mathrm{h}}\mu_0}$ is the speed of light in the host medium, and $k_{\mathrm{h}}$ the wave number in the host medium. The $\mathbf{T}^{j,j'=1,1}_{\mathrm{eff,\nu\nu',cart}}$ are block matrices building $\Teffdip$ in the Cartesian basis:

\begin{equation}
	\Teffdip\equiv\begin{pmatrix}
\mathbf{T}^{j,j'=1,1}_{\mathrm{eff,EE,cart}}&\mathbf{T}^{j,j'=1,1}_{\mathrm{eff,EM,cart}}\\
\mathbf{T}^{j,j'=1,1}_{\mathrm{eff,ME,cart}}&\mathbf{T}^{j,j'=1,1}_{\mathrm{eff,MM,cart}}\end{pmatrix}.
\end{equation}

\noindent The change from the effective dipolar T-matrix in the basis of vector spherical waves of well-defined helicity to the electric/magnetic basis in Cartesian coordinates is achieved via simple matrix multiplications [see Equation~(6) in Ref.~\onlinecite{Fernandez-Corbaton:2020}].

\noindent The internal fields in a unit volume in the homogenized lattice are therefore a sum of the incident and the depolarization fields

\begin{align}\label{eq:Inter}
\begin{pmatrix}
\bm{E}\\
\bm{H}
\end{pmatrix}
= \begin{pmatrix}
\bm{E}_{\mathrm{ext}}\\
\bm{H}_{\mathrm{ext}}
\end{pmatrix}
-
\begin{pmatrix}
\frac{1}{\varepsilon_{\mathrm{h}}}\mathbf{L}&0\\
0&\frac{1}{\mu_{\mathrm{0}}}\mathbf{L}
\end{pmatrix}
\begin{pmatrix}
\bm{P}_{\mathrm{eff,e}}\\
\bm{P}_{\mathrm{eff,m}}
\end{pmatrix},
\end{align}
where $\mathbf{L}$ is the depolarization matrix, which depends on the geometrical shape of the unit volume. The latter has the same shape as the unit cell which it has to fill. For a cube, $\mathbf{L}=(1/3)\mathbf{I}_3$, if $\Teffdip$ and the polarizations are considered at the origin of the lattice. For a cuboid, a formula for $\mathbf{L}$ can be found in \cite{doi:10.1021/jp509245u,doi:10.1063/1.367113}.
 
In frequency domain, the common bi-anisotropic constitutive relations relating the electric displacement $\bm{D}$ and magnetic flux density $\bm{B}$ to the fields inside a material consisting of the lattice read

\begin{align}\label{eq:ConsRelMeth}
	\begin{pmatrix}
		\bm{D}\\
		\bm{B}
	\end{pmatrix}
	=
	\begin{pmatrix}
	\bm{\varepsilon}_{\mathrm{eff}}&\mathrm{i}\bm{\kappa}_{\mathrm{eff}}\sqrt{\varepsilon_0\mu_0}\\
	\mathrm{i}\bm{\gamma}_{\mathrm{eff}}\sqrt{\varepsilon_0\mu_0}&\bm{\mu}_{\mathrm{eff}}
	\end{pmatrix}
	\begin{pmatrix}
	\bm{E}\\
	\bm{H}
	\end{pmatrix},
	\end{align} 
	
\noindent where $\bm{\varepsilon}_{\mathrm{eff}}$ is the effective tensorial permittivity, $\bm{\mu}_{\mathrm{eff}}$ the permeability, and $\bm{\kappa}_{\mathrm{eff}}$ and $\bm{\gamma}_{\mathrm{eff}}$ describe the coupling between the electric and magnetic fields. We aim at relating the effective material parameters to the effective polarizations and to the $\Teffdip$.

The electric displacement and the magnetic flux density can be expressed via the electric and magnetic polarizations $\bm{P}_{\mathrm{eff,e}}$ and $\bm{P}_{\mathrm{eff,m}}$ of the lattice as

 \begin{align}\label{eq:ConsRel2}
 \begin{pmatrix}
 \bm{D}\\
 \bm{B}
 \end{pmatrix}
 =
 \begin{pmatrix}
 {\varepsilon}_{\mathrm{h}}\mathbf{I}_3&0\\
 0&{\mu}_{\mathrm{0}}\mathbf{I}_3
 \end{pmatrix}
 \begin{pmatrix}
 \bm{E}\\
 \bm{H}
 \end{pmatrix}
 +\begin{pmatrix}
 \bm{P}_{\mathrm{eff,e}}\\
 \bm{P}_{\mathrm{eff,m}}
 \end{pmatrix}.
 \end{align}

Equations (\ref{eq:Pol},\ref{eq:Inter},\ref{eq:ConsRel2}) can be used to obtain an expression of ($\bm{D}$,$\bm{B}$) as a function of $\Teffdip$ and ($\bm{E},\bm{H}$) by inserting the implication of Equation~(\ref{eq:Inter}) for the externally applied fields into Equation~(\ref{eq:Pol}), and substituting the polarizations in Equation~(\ref{eq:ConsRel2}) with the resulting expressions. Comparison of the result with Equation~(\ref{eq:ConsRelMeth}) gives the material parameters

\begin{align}\label{eq:EffParam}
&\begin{pmatrix}
\bm{\varepsilon}_{\mathrm{eff}}&\mathrm{i}\bm{\kappa}_{\mathrm{eff}}\sqrt{\varepsilon_0\mu_0}\\
\mathrm{i}\bm{\gamma}_{\mathrm{eff}}\sqrt{\varepsilon_0\mu_0}&\bm{\mu}_{\mathrm{eff}}
\end{pmatrix}
=
\begin{pmatrix}
{\varepsilon}_{\mathrm{h}}\mathbf{I}_3&0\\
0&{\mu}_{0}\mathbf{I}_3
\end{pmatrix}+\nonumber\\&+n\left(\mathbf{I}_6-n\cdot q\begin{pmatrix}
\frac{1}{{\varepsilon}_{\mathrm{h}}}\mathbf{T}^{j,j'=1,1}_{\mathrm{eff,EE,cart}}\mathbf{L}&\mathrm{i}c_{\mathrm{h}}\mathbf{T}^{j,j'=1,1}_{\mathrm{eff,EM,cart}}\mathbf{L}\\
\mathrm{-i}\frac{Z_{\mathrm{h}}}{{\varepsilon}_{\mathrm{h}}}\mathbf{T}^{j,j'=1,1}_{\mathrm{eff,ME,cart}}\mathbf{L}&c_{\mathrm{h}}Z_{\mathrm{h}}\mathbf{T}^{j,j'=1,1}_{\mathrm{eff,MM,cart}}\mathbf{L}
\end{pmatrix}\right)^{-1}\times
 \\& \times q\begin{pmatrix}
\mathbf{T}^{j,j'=1,1}_{\mathrm{eff,EE,cart}}&\mathrm{i}Z_{\mathrm{h}}\mathbf{T}^{j,j'=1,1}_{\mathrm{eff,EM,cart}}\\
\mathrm{-i}Z_{\mathrm{h}}\mathbf{T}^{j,j'=1,1}_{\mathrm{eff,ME,cart}}&Z_{\mathrm{h}}^2\mathbf{T}^{j,j'=1,1}_{\mathrm{eff,MM,cart}}\end{pmatrix}\nonumber\mathrm{.}
\end{align}

The depolarization tensor $\mathbf{L}$ in Equation~(\ref{eq:EffParam}) describes the depolarization of a unit cell by the external fields but not the interaction between the scatterers inside the lattice. Such interaction is already incorporated in $\Teff$. In particular, the lattice interactions modify the dipolar terms in $\Teffdip$ and hence the effective material parameters.

\subsection{Using CELES for computing the scattering cross-section of a cluster of scatterers}
Cross-sections for the scattered waves can be computed using the software CELES \cite{EGEL2017103} as:

\begin{align}
C_{\mathrm{sca}}=\frac{2\pi\sum_{\bm{\hat{k}}_{\mathrm{sca}}}\vert E_{\bm{\hat{k}}_{\mathrm{sca}}}\vert^2}{k_{\mathrm{sca,h}}^2},
\end{align}

\noindent where $E_{\bm{\hat{k}}_{\mathrm{sca}}}$ are the scattered field amplitudes computed with CELES, $k_{\mathrm{sca,h}}$ is the absolute value of the wave vector of the scattered plane waves in the surrounding medium, and $\bm{\hat{k}}_{\mathrm{sca}}$ is the propagation direction of the scattered plane waves.

\medskip
\textbf{Acknowledgements} \par 
We are thankful to Yannick Augenstein and Aristeidis Lamprianidis for providing useful information regarding the capabilities of different Maxwell solvers. We are thankful to Prof. Dr. Eva Blasco for useful conversations about 3D laser printing. R.V., D.B., M.K., and C.R. acknowledge support by the Deutsche Forschungsgemeinschaft (DFG, German Research Foundation) under Germany’s Excellence Strategy via the Excellence Cluster 3D Matter Made to Order (EXC-2082/1-390761711), from the SFB 1173 (Project-ID No. 258734477), and from the Carl Zeiss Foundation via the CZF-Focus@HEiKA Program. M.K., C.H., and C.R. acknowledge funding by the Volkswagen Foundation. I.F.C., and C.R. acknowledge support by the Helmholtz Association via the Helmholtz program “Materials Systems Engineering” (MSE). B.Z. and C.R. acknowledge support by the KIT through the “Virtual Materials Design” (VIRTMAT) project. M.K. and C.R. acknowledge support by the state of Baden-Württemberg through bwHPC and the German Research Foundation (DFG) through grant no. INST 40/575-1 FUGG (JUSTUS 2 cluster) and the HoreKa supercomputer funded by the Ministry of Science, Research and the Arts Baden-Württemberg and by the Federal Ministry of Education and Research. We are grateful to the company JCMwave for their free provision of the FEM Maxwell solver JCMsuite.

\medskip 
\textbf{Conflict of Interest} \par
The authors declare no conflicts of interest.

\bibliography{bibliography}

\end{document}